\documentclass[aps,prd,twocolumn,superscriptaddress,nofootinbib,longbibliography]{revtex4-2}
\usepackage{amsmath,amsfonts,amssymb,mathtools}
\usepackage{bm}
\usepackage{braket}
\usepackage{physics}

\usepackage{graphicx}
\usepackage{subcaption}
\usepackage{float}
\usepackage{diagbox}

\usepackage{dcolumn}
\usepackage{array}
\usepackage{multirow}
\usepackage{makecell}

\usepackage[dvipsnames]{xcolor}
\usepackage{ulem,soul,changes}
\usepackage{verbatim}
\usepackage{cases}
\usepackage{romannum}
\usepackage{mhchem}
\usepackage{enumitem}
\usepackage{indentfirst}

\usepackage{geometry}
\usepackage[toc,title,page]{appendix}

\usepackage[colorlinks]{hyperref}
\AtBeginDocument{
	\hypersetup{
		urlcolor=black,
		citecolor=green,
		linkcolor=blue,
	}
}

\AtBeginDocument{\pagenumbering{arabic}}

\newcolumntype{M}[1]{>{\centering\arraybackslash}m{#1}}
\newcolumntype{N}{@{}m{0pt}@{}}

\begin{document}
	
	\title{Spin-Dependent Force and Inverted Harmonic Potential for Rapid Creation of Macroscopic Quantum Superpositions}
	
	\newcommand{\affone}{Centre for Quantum Computation and Communication Technology, School of Mathematics and Physics, University of Queensland, Brisbane, Queensland 4072, Australia}
	\newcommand{\afftwo}{Department of Physics and Astronomy, University College London, Gower Street, WC1E 6BT London, United Kingdom.}
	\newcommand{\affthree}{Van Swinderen Institute, University of Groningen, 9747 AG Groningen, The Netherlands.}

	\author{Run Zhou}
	\affiliation{\affthree}
	%
	
	\author{Qian Xiang}
	\affiliation{\affthree}
	
	\author{Anupam Mazumdar}
	\affiliation{\affthree}

	\date{\today}

	\begin{abstract}
		Creating macroscopic spatial superposition states is crucial for investigating matter-wave interferometry and advancing quantum sensor technology. Currently, two potential methods exist to achieve this objective. The first involves using inverted harmonic potential (IHP) to spatially delocalize quantum states through coherent inflation \cite{Romero-Isart2017}.  The second method employs a spin-dependent force to separate two massive wave packets spatially \cite{scala2013matter}. The disadvantage of the former method is the slow initial coherent inflation, while the latter is hindered by the diamagnetism of spin-embedded nanocrystals, which suppresses spatial separation. In this study, we integrate two methods: first, we use the spin-dependent force to generate initial spatial separation, and second, we use IHP to achieve coherent inflating trajectories of the wavepackets. This approach enables the attainment of massive large spatial superposition in minimal time. For instance, a spatial superposition with a mass of $10^{-15}$ kg and a size of around 50 $\mu$m is realized in $0.1$ seconds. We also calculate the evolution of wave packets in both harmonic potential (HP) and IHP using path integral approach.
	\end{abstract}
	
	\maketitle	
	\section{Introduction} 
	
	The significant interest in massive quantum superposition states primarily arises from three aspects. The first is the exploration of the quantum-classical boundary \cite{arndt:2014testing}. Quantum interference, from electrons to macromolecules ($10^{-31}-10^{-23}$ kg), has been observed in contemporary experiments, demonstrating their quantum nature \cite{TonomuraEtAl1989,kovachyQuantumSuperpositionHalfmetre2015,Arndt:1999kyb,gerlich2011quantum,FeinEtAl2019}. This leads to quests about whether quantum superposition states can also be achieved for objects of larger mass \cite{Kialka:2022vhj}. The second aspect is their utility in validating theoretical models \cite{Bose:2023gwh}. For instance, they can be employed to test wave function collapse theories \cite{bassi2013models,Bassi:2023hvn}, modified quantum mechanical frameworks \cite{AdlerBassi2009,Bassi:2003gd,PhysRevA.84.052121}, and examine the weak equivalence principle \cite{Bose:2022czr}. Additionally, they may reveal the quantum nature of gravity by combining two massive spatial superposition states \cite{bose2017spin,ICTS,marletto2017gravitationally}.
	
	The third aspect is that they can act as highly sensitive quantum sensors, detecting phenomena such as the Casimir force and dipole interactions \cite{vandeKamp:2020rqh,Schut:2023eux,Schut:2023hsy,Marshman:2023nkh}, gravitational waves \cite{Arvanitaki:2012cn,Marshman:2018upe}, quantum sensors for detecting accelerations, and inertial rotations~\cite{Toros:2020dbf,Wu:2022rdv},
	dark matter \cite{Kilian:2024fsg}, physics beyond the Standard Model~\cite{Barker:2022mdz}, testing massive graviton~\cite{Elahi:2023ozf}, non-local gravitational interaction~\cite{Marshman:2019sne,Vinckers:2023grv}, and analogue of light bending experiment in quantum gravity~\cite{Biswas:2022qto}.
	
	Advancements in quantum technology have enabled the fabrication of massive quantum superposition states. A critical challenge in realizing these states is decoherence, induced by gas molecule scattering and the emission and absorption of thermal photons, dipoles and electromagnetic interactions~\cite{Schlosshauer:2019ewh,PhysRevA.84.052121,Fragolino:2023agd,Schut:2023tce}. However, this decoherence effect can be effectively minimized through levitation mechanics in ultrahigh vacuum environments \cite{LevitationControl2021}. It is now feasible to cool either the internal degrees of freedom (phonons) or the mechanical degrees of freedom (center-of-mass (CoM) motion) of nano-objects ranging from $10^{-16}$ to $10^{-14}$ kg to a quantum ground state \cite{Teufel:2011smx,Chan:2011ivv,GieselerEtAl2012,Delic:2020ndp,Kamba:2023zoq}. Additionally, atom chips are employed to control magnetic fields precisely \cite{Keil:2016bmi}. Recently, a full-loop Stern-Gerlach interferometer for \ce{^{87}Rb} atoms was realized for the first time using magnetic fields generated by atom chips \cite{Margalit:2020qcy}. Furthermore, embedding a single nitrogen-vacancy (NV) centre in nanodiamonds has achieved electron spin coherence times of $\mathcal{O}(1)$ ms \cite{PettitEtAl2017, DelordEtAl2018}. However, by mapping the electronic spin onto a nearby \ce{^{13}C} nuclear spin, the coherence time can be significantly extended to nearly $\mathcal{O}(1)$ s.~\cite{Abobeih2018}.
	
	Based on these cutting-edge techniques, numerous experimental schemes for creating macroscopic spatial superposition states have been proposed \cite{PhysRevA.84.052121,scala2013matter,yin2013large,Bateman:2013zna,SticklerEtAl2018a,WanEtAl2016,Romero-Isart2017,pino2018chip,wood2022spin,Zhou:2022frl,Pedernales_2020,Marshman:2021wyk,Zhou:2022jug,Neumeier:2022czd,Marshman:2023nkh,Roda-Llordes:2023odc}. A natural method to achieve a delocalized quantum state is to allow a quantum wave packet to evolve freely \cite{PhysRevA.84.052121}. However, this delocalization process is slow. For instance, consider a silica microsphere with a mass of $10^{-15}$ kg trapped by a magnetic field at a frequency of 100 Hz \cite{slezakSilica2018}. Its initial wave packet spatial width is approximately $10^{-11}$ m. After 1 s of free evolution, the wave packet width becomes about $10^{-9}$ m, roughly one-thousandth of its size. To accelerate this delocalization process, an IHP can induce coherent inflation \cite{Romero-Isart2017,pino2018chip}. This approach allows the coherent length of a $10^{-14}$ kg nanoparticle to increase to around 1 $\mu$m in 0.6 s, making the coherence length comparable to its size. Another method for creating macroscopic spatial superposition states involves utilizing spin-dependent forces, such as using diamond embedded with a NV center \cite{scala2013matter,yin2013large,Marshman:2021wyk,Pedernales_2020}. Initially, a pulse is used to place the electron spin of the NV center in a superposition state, followed by applying a magnetic field to induce spatial splitting, similar to the Stern-Gerlach experiment. The advantage of this method is the ease of preparing the initial superposition state and reading out the final spin state \cite{Robledo:2011zlo}. However, the diamagnetism of the diamond suppresses the spatial separation of the wave packets when an external magnetic field is applied \cite{Zhou:2022frl,Zhou:2024pdl,Zhou:2022jug}, limiting the direct increase of the superposition size through enhanced magnetic field gradients.
	
	In this work, we combine spin-dependent forces and  IHP to achieve a massive large spatial superposition in a relatively short time. Initially, the spin-dependent force is used to create a spatial separation between two massive wave packets. Subsequently, the IHP facilitates rapid separation of the wave packets. This approach addresses the challenges of slow acceleration in the early stages of the IHP \cite{Romero-Isart2017,pino2018chip} and the suppression of superposition size due to diamagnetism \cite{Pedernales_2020,
		Marshman:2021wyk,Marshman:2023nkh,Zhou:2022frl,Zhou:2024pdl,Zhou:2022epb,
		Zhou:2022jug,Braccini:2023eyc}.
	
	The paper is organized as follows. Section \ref{experimental_scheme} presents the specific experimental protocol and the magnetic fields required to construct the HP and IHP for the experiment. Section \ref{analytic_analysis} provides an analytical solution for the classical trajectories of the wave packets at each experiment stage. In Section \ref{numerical_results}, we numerically calculate the classical trajectories of the wave packet without approximations for the nonlinear magnetic field and compare these results with the analytical solution. Section \ref{wave_packet_evolution} discusses the evolution of the wave packet under HP and IHP using path integrals. We also examine the effect of fluctuations in the magnetic field gradient and initial position on wave packet contrast for both the HP and IHP cases in this section. Finally, we conclude our findings in Section \ref{conclusion}.
	\section{Experimental scheme}\label{experimental_scheme}
	
	The Hamiltonian of the nanodiamond embedded with a NV center in the presence of an external magnetic field is given by:
	\begin{align}\label{hamiltonian1}
		\hat{\vb{H}} =\frac{1}{2 m} \hat{\vb{P}}^2 - \frac{\chi_{\rho} m }{2 \mu_0} \hat{\vb{B}}^2+\hbar \gamma_e \hat{\vb{S}}\vdot \hat{\vb{B}}+\hbar D \hat{S}_{\text{nv}}^2,
	\end{align}
	where the first term represents the kinetic energy of the nanodiamond, with $\hat{\vb{P}}$ as its momentum and $m$ as its mass. The second term signifies the magnetic energy of a diamagnetic material (nanodiamond) in a magnetic field, with $\chi_{\rho}=-6.2\times 10^{-9}\,\,\text{m}^{3}/\text{kg}$ as the mass susceptibility and $\mu_{0}$ as the vacuum permeability. The third term describes the interaction between the electron spin and the magnetic field, with $\hbar$ as the reduced Planck constant, $\gamma_{e}$ as the electronic gyromagnetic ratio, $\hat{\vb{S}}$ as the spin operator, and $\hat{\vb{B}}$ as the magnetic field. The final term represents the zero-field splitting of the NV center, where $D=(2\pi)\times2.8$ GHz and $\hat{S}_{\text{nv}}$ is the spin component operator aligned along the NV axis. Throughout this work, we assume that the nanodiamond's rotational angular momentum is negligible. For discussions on finite angular momentum and its implications, see Ref.~\cite{Zhou:2024pdl}.
	
	The experimental protocol consists of five stages:  
	\begin{figure}
		\centering
		\includegraphics[width=0.8\linewidth]{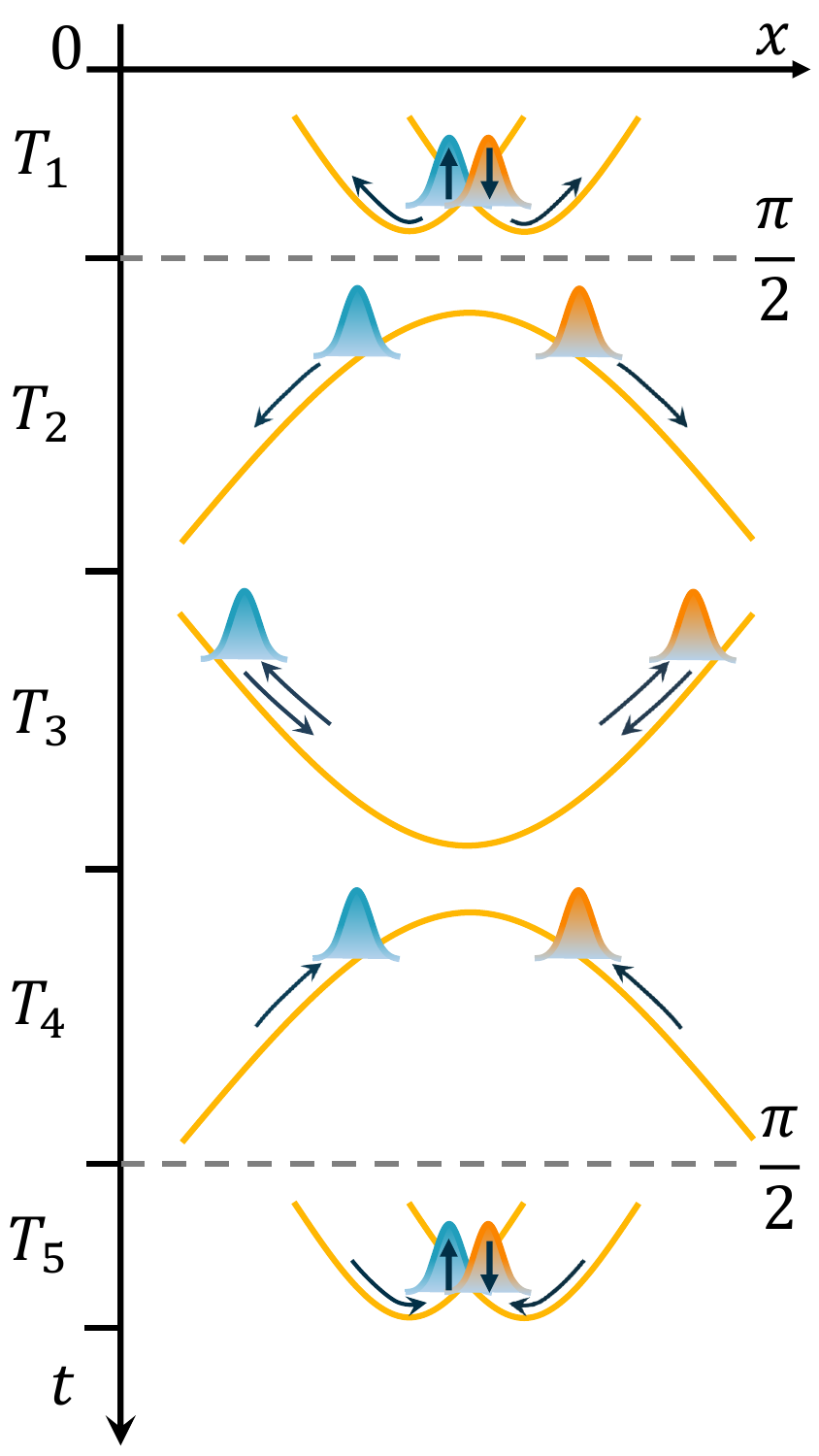}
		\caption{Schematic of the experimental scheme. The light orange quadratic curves represent harmonic or inverted harmonic potentials (HPs and IHPs). The blue and orange curves with shaded regions depict wave packets in a superposition state. The arrows inside the wave packets indicate the corresponding electron spin states (spin up or spin down). The arrows along the potential curves denote the direction of wave packet motion. The double arrows in the return stage signify the initial separation and subsequent recombination. The time axis runs from top to bottom, corresponding to Stages 1–5. The evolution times of each stage are denoted by $T_{1}$ to $T_{5}$. The gray dashed lines indicate the moments when $\pi/2$ pulses are applied.}
		\label{sketch_scheme}
	\end{figure}
	\begin{itemize}
		\item {\bf Initial state:} 
		
		The system is initially prepared in the spin superposition state $\ket{S_{x}}=(\ket{+1}+\ket{-1})/\sqrt{2}$, where $\ket{+1}$ and $\ket{-1}$ are eigenstates of the spin operator $\hat{S}_{x}$ with eigenvalues $+1$ and $-1$, respectively. We assume that the NV axis is aligned along the x-direction to prevent rotations due to misalignment with the magnetic field~\cite{Zhou:2024pdl,Japha:2022xyg}.
		
		\item {\bf Stage 1: Initial separation stage} 
		
		The HP is activated, causing the two wave packets to move in opposite directions, leading to a small spatial separation.
		
		\item {\bf Stage 2: Enhancement stage} 
		
		After half a period in the HP, the wave packets reach their maximum initial separation. At this point, the HP is turned off and the IHP is activated. A $\pi/2$ pulse is then applied to transform the state $\ket{S_{x}}=(\ket{+1}+\ket{-1})/\sqrt{2}$ into $\ket{S_{x}}=\ket{0}$\footnote{This transformation is not strictly necessary but simplifies the equations of motion and ensures trajectory symmetry. Even without this step, the IHP can still be used to enlarge the superposition size and recombine the wave packets by fine-tuning the magnetic field.}~\cite{TaylorEtAl2008,LevineEtAl2019}. The pulse duration is considered negligible. Subsequently, the wave packets rapidly separate in the IHP, achieving a large spatial superposition within a short time\footnote{The evolution time depends on the desired superposition size: longer times yield greater separations.}.
		
		\item {\bf Stage 3: Return stage} 
		
		The IHP is turned off, and the HP is reactivated. The wave packet velocities decrease, eventually reversing direction. The maximum spatial separation occurs at this stage.
		
		\item {\bf Stage 4: Deceleration stage} 
		
		When the wave packet velocities reach the same magnitude (but in the opposite direction) as at the end of Stage 2\footnote{The goal of this stage is to reverse the wave packet velocity. Exact velocity matching is not strictly required but ensures trajectory symmetry. If the velocities differ, trajectory closure can still be achieved by fine-tuning the magnetic field.}, the HP is turned off and the IHP is turned on again. The wave packets gradually decelerate.
		
		\item {\bf Stage 5: Recombination stage} 
		
		The IHP is turned off, and a $\pi/2$ pulse is applied to restore the spin state to $\ket{S_{x}}=(\ket{+1}+\ket{-1})/\sqrt{2}$. The HP is then activated, and the magnetic field gradient is fine-tuned to ensure that the two wave packet trajectories recombine after half a period.
	\end{itemize}
	This sequence is designed to achieve a large spatial superposition within a short time (approximately 0.1 s) and to ensure the recombination of the center-of-mass (CoM) trajectories, thereby restoring spin coherence~\cite{Margalit:2020qcy}. Figure~\ref{sketch_scheme} provides an overview of these five stages. To maintain clarity, we use terms such as “Initial Separation Stage” and “Stage 1” interchangeably in subsequent discussions.
	
	For simplicity, we consider only the x-component of the magnetic field to construct the one-dimensional HP and IHP. In realistic systems, magnetic fields must satisfy Maxwell’s equations and cannot be strictly one-dimensional. However, by applying a bias field along the x-direction and a constraining field along the y-direction, the motion of the nanodiamond can be effectively restricted to one dimension. A detailed analysis is provided in Appendix~\ref{one_dimension_motion}.
	
	The HP is generated using a linear magnetic field~\cite{scala2013matter,Pedernales_2020}:
	\begin{align}\label{linear_magnetic_field}
		\hat{B}_x = B_{0} + \eta_{l} \hat{x},
	\end{align}
	where $B_{0}$ is the bias field along x and $\eta_{l}$ is the field gradient. Substituting Eq.(\ref{linear_magnetic_field}) into the Hamiltonian Eq.(\ref{hamiltonian1}) one have:
	\begin{align}\label{hamiltonian2}
		\hat{H}_{x}^{\text{H}} = &\frac{1}{2 m} \hat{P}_x^2 + \frac{1}{2}m\omega_{h}^{2} \hat{x}^2 +\qty(\hat{S}_x \hbar \gamma_e \eta_l - \frac{\chi_{\rho} m }{ \mu_0}B_{0}\eta_{l})\hat{x} \nonumber\\
		&-\frac{\chi_{\rho} m }{2 \mu_0}B_{0}^{2} + \hat{S}_x \hbar \gamma_e B_{0} + \hbar D \hat{S}_{\text{nv}}^2.
	\end{align}
	The superscript ``H'' denotes the Hamiltonian associated with the HP. The corresponding frequency is given by  
	\begin{align}
		\omega_{h} = \sqrt{-\frac{\chi_\rho}{\mu_{0}}} \eta_{l}.
	\end{align}
	
	Similarly, the IHP is generated using a nonlinear magnetic field~\cite{Zhou:2022frl}:
	\begin{align}\label{nonlinear_magnetic_field}
		\hat{B}_x = B_{0} - \eta_{n} \hat{x}^{2},
	\end{align}
	where $\eta_{n}$ is the gradient parameter characterizing the nonlinear field, with units of ${\rm T/m^{2}}$. Substituting Eq.~(\ref{nonlinear_magnetic_field}) into the Hamiltonian Eq.~(\ref{hamiltonian1}), we obtain:
	\begin{align}\label{hamiltonian3}
		\hat{H}_x = &\frac{1}{2 m} \hat{P}_x^2+\left(\frac{\chi_\rho m}{\mu_0} B_0 \eta_n-\frac{\chi_\rho m}{2 \mu_0} \eta_n^2 \hat{x}^2\right) \hat{x}^2\nonumber\\
		&-\frac{\chi_\rho m}{2 \mu_0} B_0^2+\hbar D \hat{S}_{\text{nv}}^2.
	\end{align}
	Here, we have used the fact that $S_{x}\vdot B_{x} = 0$ in the IHP stages, as the spin state is set to $S_{x}=\ket{0}$. From Eq.~(\ref{hamiltonian1}), the IHP is effectively realized under the condition\footnote{If this condition is not met, the resulting potential takes a quartic form, which can also be employed to generate macroscopic spatial superpositions~\cite{Roda-Llordes:2023odc}. The wave packet dynamics in quartic potentials can be solved both analytically and numerically~\cite{Riera-Campeny:2023xhw,Roda-Llordes:2023dwm}.}:
	\begin{align}
		\abs{\expval{\hat{x}}} \ll \sqrt{\frac{2B_{0}}{\eta_{n}}}.
	\end{align} 
	Under this condition, the Hamiltonian simplifies to:
	\begin{align}\label{hamiltonian4}
		\hat{H}_x^{\text{I}} = \frac{1}{2 m} \hat{P}_x^2 - \frac{1}{2}m\omega_{r}^{2}\hat{x}^2 - \frac{\chi_\rho m}{2 \mu_0} B_0^2+\hbar D \hat{S}_{\text{nv}}^2.
	\end{align}
	The superscript ``I'' denotes the Hamiltonian corresponding to the IHP, with the associated frequency given by:
	\begin{align}
		\omega_{r} = \sqrt{-\frac{2\chi_\rho B_0 \eta_n}{\mu_0}}.
	\end{align}
	\section{C\lowercase{o}M trajectory and superposition size}\label{analytic_analysis}
	The expectation value of position operator $\hat{x}$ satisfies the equation of motion:
	\begin{align}\label{Ehrenfest_equation}
		\dv{\langle\hat{x}\rangle}{t} &= \frac{i}{\hbar}\langle[\hat{H}, \hat{x}]\rangle.
	\end{align}
	Since the trajectories of the two wave packets are completely symmetric, to simplify the calculation process, we take the wave packet with the spin quantum number $S_{x}=1$ as an example to calculate the classical trajectory. For convenience of representation, we make the conventions shown in Table \ref{symbol_meaning}.
	\begin{table}[t]
		\setlength{\tabcolsep}{0.5\tabcolsep}
		\renewcommand{\arraystretch}{2.2}
		\begin{tabular}{c c}
			\hline
			\textbf{Symbol}                 & \textbf{Meaning}                                                                                       \\ \hline
			$\omega_{i}$           & \makecell{The frequency of the harmonic or \\ inverted harmonic potential at the i-th stage.} \\ 
			$\expval{\hat{x}}_{i}$ & The classical trajectory of the i-th stage.                                                   \\ 
			$X_{i}$                & \makecell{The classical position at the end of \\ the i-th stage.}                                          \\ 
			$\dot{X}_{i}$          & \makecell{The classical velocity at the end of \\ the i-th stage.}                                          \\ 
			$t_{i}$                & The time variable at the i-th stage.                                                         \\ 
			$T_{i}$                & The time interval at the i-th stage.\footnote{For example, at the beginning of the i-th stage $t_{i}=0$ and at the end of that stage $t_{i}=T_{i}$.}                                                          \\ \hline
		\end{tabular}
		\caption{The mathematical symbols that appear in calculating classical trajectories and their physical interpretations.}
		\label{symbol_meaning}
	\end{table}
	
	{\bf Stage 1} --- Substituting the Hamiltonian Eq.(\ref{hamiltonian2}) into Eq.(\ref{Ehrenfest_equation}) and then taking the second order derivative of the expectation value of the position operator with respect to time gives:
	\begin{align}\label{second_derivative_of_x}
		\dv[2]{\expval{\hat{x}}_{1}}{t} = -\omega_{1}^{2}\expval{\hat{x}}_{1} - \frac{\hbar \gamma_e \eta_l}{m} + \frac{\chi_{\rho}}{\mu_{0}}B_{0}\eta_{l},
	\end{align}
	where $\omega_{1}$ is the frequency of the HP in the initial separation stage. The $B_{0}$ term in Eq.(\ref{second_derivative_of_x}) does not affect the maximum superposition size in the initial separation stage. To be consistent with the coordinates of the later stages, we set $B_{0}$ in the initial separation stage equal to zero. Considering the initial conditions $\expval{\hat{x}(0)} = 0$ and $\langle\dot{\hat{x}}(0)\rangle=0$, the solution of Eq.(\ref{second_derivative_of_x}) is:
	\begin{align}\label{EoMstage1}
		\expval{\hat{x}}_{1}=\frac{\hbar \gamma_e\eta_{l}}{\omega_{1}^{2}m}(\cos(\omega_{1} t_{1}) - 1 ).
	\end{align}
	When $t_{1}=\pi/\omega_{1}$, the superposition size achieves the maximum value $4\hbar \gamma_e\mu_{0}/\chi_{\rho}m\eta_{l}$ in the initial separation stage. The position of the CoM at this point is taken as the initial condition to solve the equation of motion for the enhancement stage. 
	
	{\bf Stage 2} --- Using Eq.(\ref{Ehrenfest_equation}) again and considering the Hamiltonian in Eq.(\ref{hamiltonian4}), one can obtain the CoM trajectory for the enhancement stage:
	\begin{align}\label{EoMstage2}
		\expval{\hat{x}}_{2} = X_{1}\cosh (\omega_{2} t_{2}),
	\end{align}
	where 
	\begin{align}
		X_{1} = - 2\hbar \gamma_e\eta_{l}/\omega_{1}^{2}m.
	\end{align} 
	The $\omega_{2}$ is the frequency of the IHP in the enhancement stage.
	
	{\bf Stage 3} --- At the end of the enhancement stage, the position and velocity of the CoM are:
	\begin{align}
		X_{2}&=X_{1}\cosh (\omega_{2} T_{2}),\nonumber\\
		\dot{X}_{2}&=X_{1}\omega_{2}\sinh (\omega_{2}T_{2}).
	\end{align}  
	Taking the position and velocity of the CoM as the initial conditions and then combining Eq.(\ref{hamiltonian2}) and (\ref{Ehrenfest_equation}) yields the trajectory of the CoM in the return stage:
	\begin{align}\label{equation_of_motion_in_stage3}
		\expval{\hat{x}}_{3}&=X_{2}\cos(\omega_{3} t_{3}) + \frac{\dot{X}_{2}}{\omega_{3}}\sin (\omega_{3} t_{3}),\nonumber\\
		&=\sqrt{X_{2}^{2}+(\dot{X}_{2}/\omega_{3})^{2}}\sin(\omega_{3}t_{3}+\phi),
	\end{align}
	where
	\begin{align}
		\phi=\arcsin\qty(\frac{X_{2}}{\sqrt{X_{2}^{2}+(\dot{X}_{2}/\omega_{3})^{2}}}).
	\end{align}
	The $\omega_{3}$ is the frequency of the HP in the return stage. The superposition size reaches its maximum value when $\sin(\omega_{3}t_{3}+\phi)=1$. The maximum superposition size is:
	\begin{align}\label{largest_SS}
		&\Delta X_{max} = \frac{T_{1}}{m}\frac{4\hbar \gamma_{e}}{\pi}\sqrt{\frac{\mu_{0}}{-\chi_{\rho}}}\zeta,
	\end{align}
	where
	\begin{equation}
		\zeta = \sqrt{\cosh^{2}(\omega_{2}T_{2})+\qty(\frac{\omega_{2}}{\omega_{3}})^{2}\sinh^{2}(\omega_{2}T_{2})},
	\end{equation}
	is a dimensionless quantity. The maximum superposition size can be rewritten as:
	\begin{align}
		\Delta X_{max}\approx \qty(\frac{3.4\times 10^{-16}\, \text{kg}}{m})\qty(\frac{T_{1}}{\text{1 sec}})\zeta\times 10^{-6}\,\text{m}.
	\end{align}
	The time corresponding to the maximum superposition size at the stage 3 is:
	\begin{align}\label{time_for_maximum_SS}
		T^{*} = \frac{1}{\omega_{3}}\qty(\frac{\pi}{2}-\phi).
	\end{align}
	We set the time interval of stage 3 to be $T_{3}=2T^{*}$. This is not necessary, but doing so makes the enhancement and deceleration stages symmetric. This is because, at the end of stage 3, the wave packet returns to its initial position, at which point its velocity is equal in magnitude and opposite in direction to the velocity at the beginning of the stage. We can use the same IHP as in the enhancement stage to decelerate the wave packet, thereby finally closing the wave packet trajectories. At the end of stage 3, the position and velocity of the CoM are:
	\begin{align}
		X_{3}&=\sqrt{X_{2}^{2}+(\dot{X}_{2}/\omega_{3})^{2}}\sin(\omega_{3}T_{3}+\phi),\nonumber\\ \dot{X}_{3}&=\sqrt{X_{2}^{2}+(\dot{X}_{2}/\omega_{3})^{2}}\omega_{3}\cos(\omega_{3}T_{3}+\phi). 
	\end{align} 
	
	\begin{figure*}[htp]
		\centering
		\includegraphics[width=\textwidth]{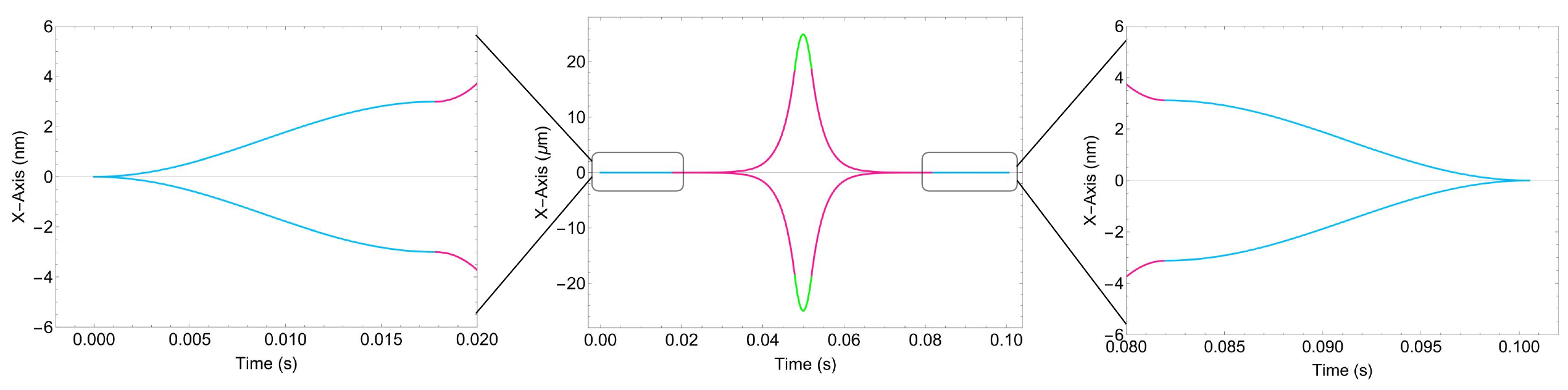}
		\caption{The numerical calculation of trajectories. The middle plot shows the complete numerical trajectories. From left to the right they are the initial separation stage (blue), the enhancement stage (purple), the return stage (green), the deceleration stage (purple), and the recombination stage (blue). On the left and right are enlarged plots of the initial separation stage and the recombination stage, respectively. The mass $m=10^{-15}$ kg. The values of other parameters at different stages are shown in Table.\ref{values_of_parameters}.}
		\label{numerical_trajectories}
	\end{figure*}
	
	{\bf Stage 4} --- Using the position and velocity at the end of the return stage as the initial conditions for the deceleration stage, similar to the enhancement stage, the CoM trajectory for the deceleration stage can be found by using Eq.(\ref{hamiltonian4}) and (\ref{Ehrenfest_equation}) as:
	\begin{align}\label{EoM_for_deceleration_stage}
		\expval{\hat{x}}_{4} = X_{3}\cosh(\omega_{4}t_{4}) + \frac{\dot{X}_{3}}{\omega_{4}}\sinh(\omega_{4}t_{4}),
	\end{align}
	where $\omega_{4}$ is the frequency of the IHP in the deceleration stage. Deriving Eq.(\ref{EoM_for_deceleration_stage}) with respect to time and making it equal to zero gives the time
	\begin{align}
		t_{4}=\frac{1}{2\omega_{4}}\ln(\frac{X_{3}\omega_{4} - \dot{X}_{3}}{X_{3}\omega_{4} + \dot{X}_{3}}), 
	\end{align}
	required for the CoM velocity to decrease to zero. Substituting the evolution time $t_{4}$ into Eq.(\ref{EoM_for_deceleration_stage}) gives the position of the CoM at this time
	\begin{align}
		X_{4} = \frac{1}{\omega_{4}}\sqrt{X_{3}^{2}\omega_{4}^{2} - \dot{X}_{3}^{2}}.
	\end{align}  
	If $X_{4}=0$ is assumed, then one have $\omega_{4} = -\dot{X}_{3}/X_{3}$. The reason for the negative sign is that the position is opposite in sign to the velocity at the end of return stage and $\omega_{4}$ should be greater than zero. However, Substituting this $\omega_{4}$ into $t_{4}$ gives $t_{4} \rightarrow \infty$. This is because as the CoM gets closer to the origin position, the velocity
	gets smaller, and at the same time the acceleration also gets smaller and
	eventually tends to zero. Therefore the time for the CoM to decelerate to zero tends to infinity. To avoid this situation, $X_{4}$ can only take a small value other than zero. This is why the recombination stage is needed to close the CoM trajectory.
	
	{\bf Stage 5} --- The equation of motion for the final stage (recombination stage) is the same as the initial separation stage but with a different frequency. The solution is:
	\begin{align}\label{EoM_for_recombination_stage}
		\expval{\hat{x}}_{5}&=- \frac{\hbar \gamma_e\eta_{l}}{\omega_{5}^{2}m}(\cos(\omega_{5} t_{5}) + 1),\nonumber\\
		&=\frac{1}{2}X_{4}(\cos(\omega_{5} t_{5}) + 1),
	\end{align}
	where $\omega_{5}$ is the frequency of the HP in the recombination stage. The second equation in Eq.(\ref{EoM_for_recombination_stage}) holds because the CoM is required to return to the origin after half a period of motion. At this point, the position and momentum of the CoM coincide.

	
	\section{Comparing analytic and numerical results}\label{numerical_results}
	In the analytic calculation of the classical trajectories of the wave packet in Sec.\ref{analytic_analysis}, we used approximate Hamiltonian (see Eq.(\ref{hamiltonian4})) in the second and fourth stages in the presence of a nonlinear magnetic field. In the numerical calculations of this section, we use the Hamiltonian without approximation (see Eq.(\ref{hamiltonian3})).

	\begin{table}[t]
		\setlength{\tabcolsep}{-0.1\tabcolsep}
		\renewcommand{\arraystretch}{1.5}
		\begin{tabular}{c|cccc}\hline
			\backslashbox{\textbf{Stages}}{\textbf{Param.}}
			&\makebox[1.1cm]{{$B_{0}$}(T)}&\makebox[1.5cm]{$\eta_{l}$(T/m)}&\makebox[1.5cm]{$\eta_{n}$(T/$\text{m}^{2}$)}&\makebox[1.5cm]{$T_{i}$(s)}
			\\\hline
			1 & 0  & 2507             & ---              & 0.01784 \\
			2 & 10 & ---              & $1\times 10^{6}$ & 0.03000 \\
			3 & 0  & $5\times 10^{3}$ & ---              & 0.00415 \\
			4 & 10 & ---              & $ 992199.56$     & 0.03000 \\
			5 & 0  & $ 2414.07$       & ---              & 0.01853
		\end{tabular}
		\caption{The values of the parameters at each stage in the calculation of the numerical trajectories. Stages 1, 3, and 5 are HPs, so $\eta_{n}$ takes no value. At these stages, $B_{0}$ takes the value 0 in order to unify the coordinate representation of each stage, but the value of $B_{0}\neq 0 $, i.e. can not vanish. Stages 2 and 4 are IHPs, so $\eta_{l}$ does not take a value.}
		\label{values_of_parameters}
	\end{table}
	
	\begin{figure}[htp]
		\centering
		\includegraphics[width=\linewidth]{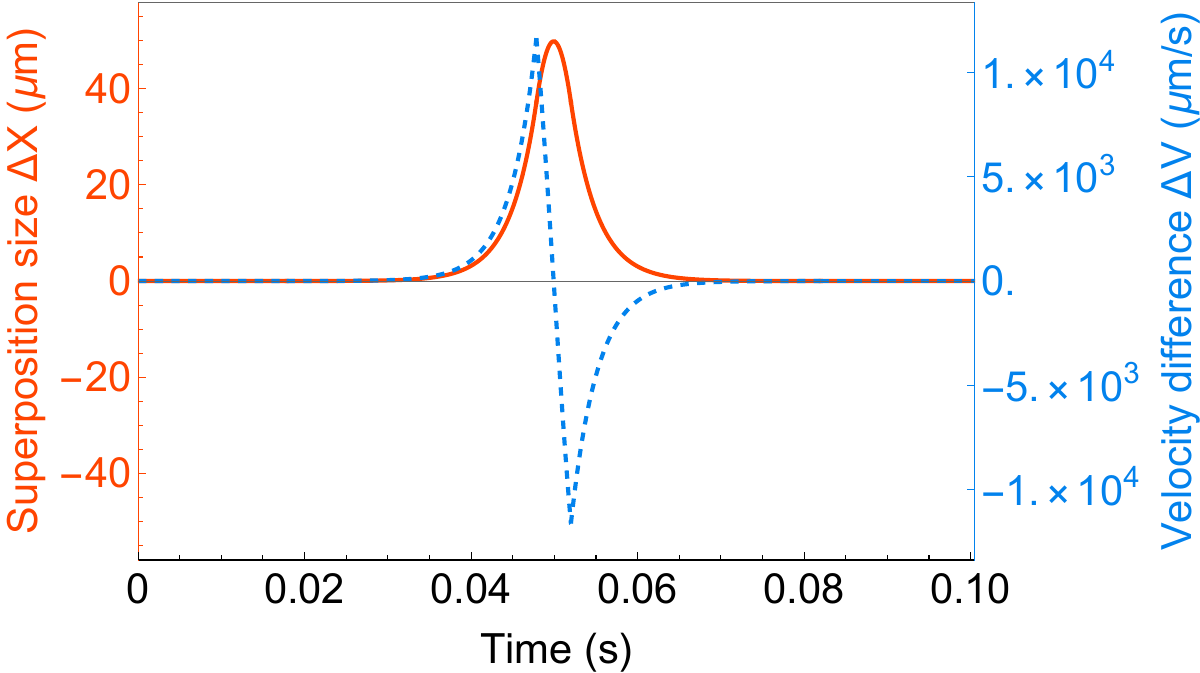}
		\caption{The superposition size and velocity differences change with time. The red solid line is the superposition size. The blue dashed line is the velocity difference. The mass $m=10^{-15}$ kg. The values of other parameters at different stages are shown in Table.\ref{values_of_parameters}.}
		\label{superposition_size_and_velocity_difference}
	\end{figure}
	
	The numerical calculation results are shown in Fig.\ref{numerical_trajectories}. The first stage is the initial separation stage. The time of this stage is $\pi/\omega_{1}$ (half a period), which is about 0.018 s. Due to their different spin states, the two wave packets move in two different HPs and yield a spatial separation of about 6 nm. The second stage is the enhancement stage. The longer the duration of this stage, the larger the superposition size obtained. However, in order to keep the total running time around 0.1 s, we set the evolution time of this stage to be 0.03 s. With the parameters in Table.\ref{values_of_parameters}, the spatial separation between the two wave packets at the end of the enhancement stage is about 37.14 $\mu$m. The third stage is the return stage. In this stage, the two wave packets move away from each other in the HP, and the speed of the wave packets decreases gradually. When the velocity decreases to 0, the spatial separation between them reaches a maximum value of about 50 $\mu$m. Then, the velocities of the two wave packets are reversed, and they gradually come closer together. We bring the wave packets back to roughly the initial position of this stage by fine-tuning the evolution time \footnote{The fine-tuning of the time is done here to make the trajectory more symmetric, but this tuning is not mandatory. We can also set different evolution times in this stage and then close the trajectory by adjusting the parameters of the later stages and keeping the total evolution time around 0.1 s.} to 0.00415 s.  The fourth stage is the deceleration stage. As analyzed in stage 4 of Sec.\ref{analytic_analysis}, this deceleration process takes an infinite time if we want the velocity to decrease to 0 when the trajectories close. So, we make the velocity decrease to 0 when the spatial separation between the wave packets is about 6 nm by fine-tuning the magnetic field gradient. The parameter values used in this stage are shown in Table.\ref{values_of_parameters} and the evolution time is 0.03 s \footnote{Here, we set the evolution time to 0.03 s to be symmetric with the second stage. We can also fix the magnetic field gradient and then fine-tune the time to make the velocity of the wave packets decrease to 0 when they are separated by about 6 nm.}. The fifth stage is the recombination stage. This stage is the inverse process of the initial separation stage. By fine-tuning the magnetic field gradient and evolution time in this stage, with parameters taking the values shown in Table.\ref{values_of_parameters}, the wave packets are able to return to the initial position and with a velocity of zero.

\begin{figure}[htp]
		\centering
		\includegraphics[width=\linewidth]{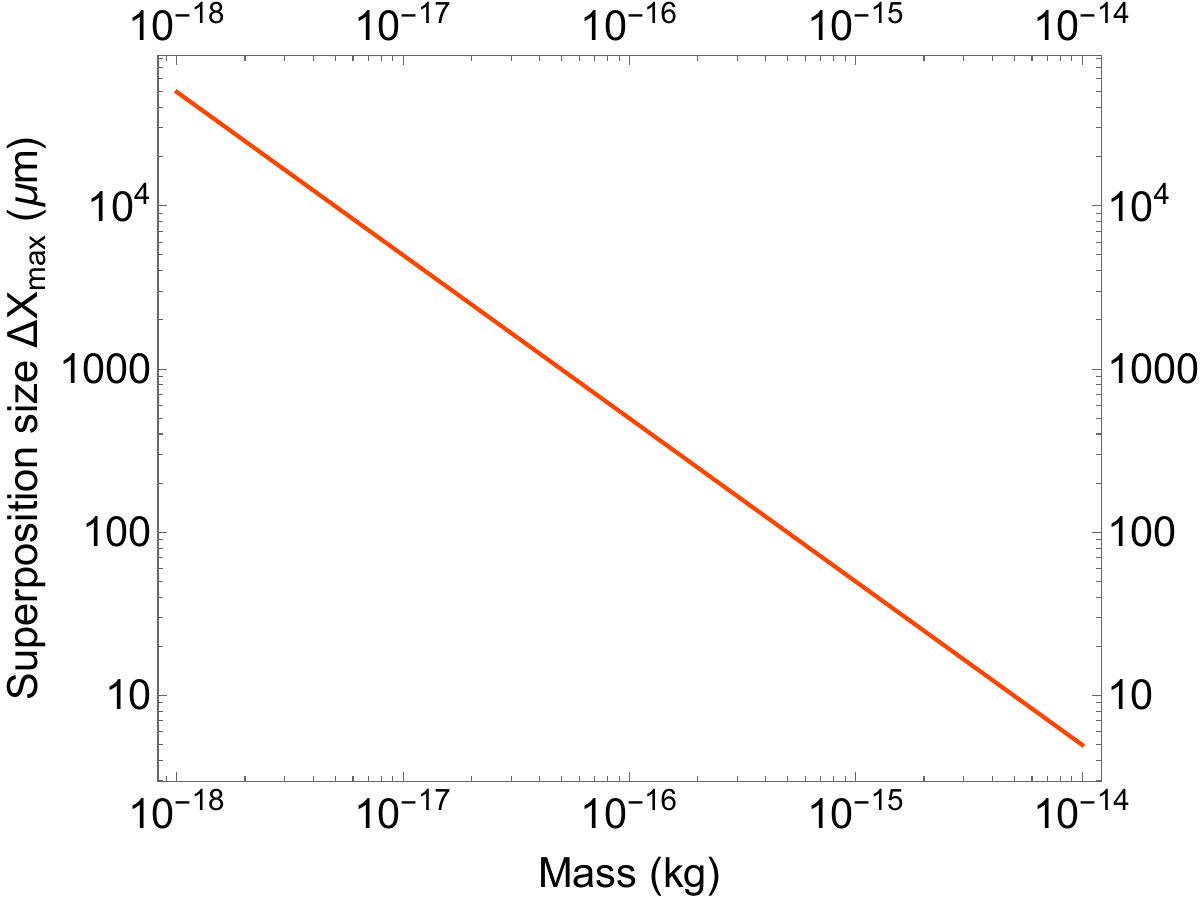}
		\caption{The scaling behaviour of the superposition size with respect to the mass. The values of other parameters at different stages are shown in Table.\ref{values_of_parameters}.}
		\label{superposition_size_scaling_behavior}
	\end{figure}

	Combining these five stages, the variations in superposition size and velocity differences are shown in Fig.\ref{superposition_size_and_velocity_difference}. The superposition size increases and then decreases, reaching a maximum value of 49.8294 $\mu$m at t = 0.0499 s. At this point, the velocity difference is $0$, and the velocity of the wave packet starts to reverse. For comparison with the analytical expression, we substitute the parameter values from Table.\ref{values_of_parameters} into Eq.(\ref{largest_SS}), which gives a maximum superposition state size of 49.8298 $\mu$m. The per cent error between the theoretical and numerical calculations is less than $0.001$\%. This indicates that the IHP approximation we made is reasonable. By fixing the magnetic field gradient and evolution time from the first to the third stage to the values shown in Table.\ref{values_of_parameters}, and then varying the mass, we obtain the scalar behaviour of the superposition size with respect to the mass, as shown in Fig.\ref{superposition_size_scaling_behavior}.

	\section{Wave packet evolution}\label{wave_packet_evolution}
	If the initial state is a Gaussian shape wave packet\footnote{The difference between Gaussian wave packet and Gaussian shape wave packet is that Gaussian wave packet maintains minimum uncertainty but Gaussian shape wave packets do not necessarily maintain this property \cite{Rauh2016}.} (GSWP), then the evolution of the wave packet under HP and IHP can be solved analytically \cite{Barton:1984ey,Yuce:2021bkt,Rajeev:2017uwk,Rauh2016}. In Appendix \ref{path_integral}, we give a detailed procedure for calculating the wave packet evolution using the path integral. In this section, we provide the main results. The general form of a GSWP can be written as: 
	\begin{small}
	\begin{align}\label{initial_wave_function}
		\psi(x,t=0) = N_{0}\exp[-\frac{(x-x_{0})^{2}}{4\sigma_{0}^{2}}
		+i\qty(\frac{a_{0}}{4}x^{2}+b_{0}x+c_{0})
		],
	\end{align}
    \end{small}
	where $N_{0}$ is the normalization factor, $\sigma_{0}$ is the wave packet width, $x_{0}$ is the center position of the wave packet, and $a_{0}$, $b_{0}$ and $c_{0}$ are the phase related factors. The factor 1/4 in front of the parameter $a_{0}$ is set for the convenience of later calculations. A GSWP remains a GSWP after it has evolved in HP and IHP. Their general solutions can be written as:
	\begin{widetext}
		\begin{align}\label{general_solutions}
			\psi(x,t)=N(t)\exp[-\frac{1}{4\sigma_{x}^{2}(t)}\qty(x-x_{c}(t))^{2}+i\qty(\frac{a(t)}{4}x^{2}+b(t)x+c(t))].
		\end{align}
	\end{widetext}
	Where $\sigma_{x}$ is the spatial width of the wave packet evolving in time and $x_{c}(t)$ is the classical equation of motion of the wave packet. For the HP case, we have:
	\begin{align}
		\sigma_{x}^{\text{H}}(t)&=\sigma_{0}\qty(\frac{\hbar^{2}}{4m^{2}\omega^{2}\sigma_{0}^{4}}\sin^{2}(\omega t)+\alpha^{2}(t))^{\frac{1}{2}},\label{width_in_HP}\\
		x_{c}^{\text{H}}(t)&=\frac{\hbar a_{0} x_{0}}{2m\omega}\sin(\omega t)+x_{0}\cos(\omega t)+\frac{\hbar b_{0} }{m\omega}\sin(\omega t),
	\end{align}
	where
	\begin{align}
		\alpha(t) = \frac{\hbar a_{0} }{2m\omega}\sin(\omega t)+\cos(\omega t)
	\end{align}
	The index ``H" indicates the expression of the physical quantity in the case of HP. For the IHP case, we have:
	\begin{align}
		\sigma_{x}^{\text{I}}(t)&=\sigma_{0}\qty(\frac{\hbar^{2}}{4m^{2}\omega^{2}\sigma_{0}^{4}}\sinh^{2}(\omega t)+\beta^{2}(t))^{\frac{1}{2}},\\
		x_{c}^{\text{I}}(t)&=\frac{\hbar a_{0} x_{0}}{2m\omega}\sinh(\omega t)+x_{0}\cosh(\omega t)+\frac{\hbar b_{0} }{m\omega}\sinh(\omega t),
	\end{align}
	where
	\begin{align}\label{beta_in_IHP}
		\beta(t)=\frac{\hbar a_{0} }{2m\omega}\sinh(\omega t)+\cosh(\omega t)
	\end{align}
	The index ``I" indicates the expression of the physical quantity in the case of IHP. The expressions for the other parameters in Eq.(\ref{general_solutions}) for the HP and IHP cases are given in Appendix \ref{path_integral}. The form of the solution of the wave packet in the IHP are the same as in the HP case, but with the replacement of $``\sin"$ with $``\sinh"$ and $``\cos"$ with $``\cosh"$. If the initial state is a Gaussian wave packet, the values of the parameters in Eq.(\ref{initial_wave_function}) are:
	\begin{align}\label{initial_conditions_for_GWP}
		N_{0}=\frac{1}{2\pi\sigma_{0}^{2}},\, a_{0}=0,\, b_{0}=\frac{p_{0}}{\hbar}\,\, \text{and}\,\, c_{0}=-\frac{p_{0}x_{0}}{\hbar},
	\end{align} 
	where $P_{0}$ is the initial momentum. Substituting these parameters into Eqs.(\ref{width_in_HP}) - (\ref{beta_in_IHP}) gives:
	\begin{align}
		\sigma_{x}^{\text{H}}(t)&=\sigma_{0}\qty(\frac{\hbar^{2}}{4m^{2}\omega^{2}\sigma_{0}^{4}}\sin^{2}(\omega t)+\cos^{2}(\omega t))^{\frac{1}{2}},\label{width_in_HP2}\\
		x_{c}^{\text{H}}(t)&=x_{0}\cos(\omega t)+\frac{p_{0} }{m\omega}\sin(\omega t),\label{EoM_in_HP}
	\end{align}
	and
	\begin{align}
		\sigma_{x}^{\text{I}}(t)&=\sigma_{0}\qty(\frac{\hbar^{2}}{4m^{2}\omega^{2}\sigma_{0}^{4}}\sinh^{2}(\omega t)+\cosh^{2}(\omega t))^{\frac{1}{2}},\label{width_in_IHP2}\\
		x_{c}^{\text{I}}(t)&=x_{0}\cosh(\omega t)+\frac{p_{0} }{m\omega}\sinh(\omega t).\label{EoM_in_IHP}
	\end{align}
	The Eqs.(\ref{width_in_HP2}) and (\ref{width_in_IHP2}) for the evolution of the spatial width of the wave packet in HP and IHP are the same as those in \cite{Rauh2016}. Eqs.(\ref{EoM_in_HP}) and (\ref{EoM_in_IHP}) are the same as the classical equations of motion (\ref{equation_of_motion_in_stage3}) and (\ref{EoM_for_deceleration_stage}).

	\subsection{Fluctuation in magnetic field and wave packet contrast}\label{fluctuation}
	
	In this section, we analyze the effect of magnetic field fluctuations on wave packet contrast. These fluctuations lead to deviations in the classical position and momentum of the wave packet, preventing the two wave packets from fully overlapping and reducing their contrast. This phenomenon is also known as the Humpty-Dumpty problem in the Stern-Gerlach interferometer \cite{Schwinger1988}.
	
	We denote the wave packets in the two arms of the interferometer as $\psi_{\text{L}}(x,t)$ and $\psi_{\text{R}}(x,t)$. The wave packet contrast is defined as:  
	\begin{align}\label{contrast1}
		C(t) := \int dx \psi_{\text{L}}^{*}(x,t) \psi_{\text{R}}(x,t).
	\end{align}
	Since we focus on the effect of classical position and momentum deviations on contrast, all parameters in $\psi_{\text{L}}(x,t)$ and $\psi_{\text{R}}(x,t)$ remain the same except for $x_{c}(t)$ (classical position) and $b(t)$ (related to classical momentum). The integral in Eq.~(\ref{contrast1}) can be expressed as:  
	\begin{align}\label{contrast}
		C(t) = \exp\left[-\frac{\Delta x^{2}}{8\sigma_{x}^{2}(t)} - \frac{\sigma_{x}^{2}(t) \Delta b^{2}}{2}\right],
	\end{align}
	where  
	\begin{align}\label{deviation1}
		\Delta x &= x_{\text{R}} - x_{\text{L}},\nonumber\\
		\Delta b &= b_{\text{R}} - b_{\text{L}}.
	\end{align}
	
	Here, $x_{\text{L}}$ and $x_{\text{R}}$ are the classical positions of the wave packets in the left and right interferometer arms, respectively. Similarly, $b_{\text{L}}$ and $b_{\text{R}}$ represent the values of parameter $b$ for the left and right wave packets. In Eq.~(\ref{contrast}), only the exponential decay terms associated with classical position and momentum deviations are retained, while amplitude and phase factors are neglected \cite{Japha2021}.  
	
	Since both the classical trajectory $x(\eta,t)$ and parameter $b(\eta,t)$ depend on the gradient $\eta$\footnote{For a linear magnetic field, $\eta$ corresponds to $\eta_{l}$. For a nonlinear magnetic field, $\eta$ corresponds to $\eta_{n}$.} and time $t$, a small fluctuation $\delta\eta$ in gradient modifies Eq.~(\ref{deviation1}) as:  
	\begin{align}\label{deviation2}
		\Delta x &= x(\eta+\delta\eta,t) - x(\eta,t),\nonumber\\
		\Delta b &= b(p(\eta+\delta\eta),t) - b(p(\eta),t).
	\end{align}
	The parameter $b$ is related to classical momentum variations, so it can be expressed in terms of momentum $p$. By substituting the expressions for $x(\eta,t)$ and $b(\eta,t)$ into Eq.~(\ref{deviation2}) for both HP and IHP cases, and then combining the results with Eq.~(\ref{contrast}), we can determine the effect of gradient fluctuations on contrast in both scenarios.  
	
	In our experimental setup, both linear and nonlinear magnetic fields are used. We analyze their gradient fluctuations separately. First, we assume a small fluctuation $\delta\eta_{l}$ in the linear magnetic field gradient while keeping the nonlinear gradient constant. By combining the equations of motion from stage 1 to stage 5, Eqs.~(\ref{EoMstage1}), (\ref{EoMstage2}), (\ref{equation_of_motion_in_stage3}), (\ref{EoM_for_deceleration_stage}), and (\ref{EoM_for_recombination_stage}), we compute the position and momentum deviations induced by this gradient fluctuation at each stage.  
	
\begin{figure}[htp]
		\centering
		\includegraphics[width=\linewidth]{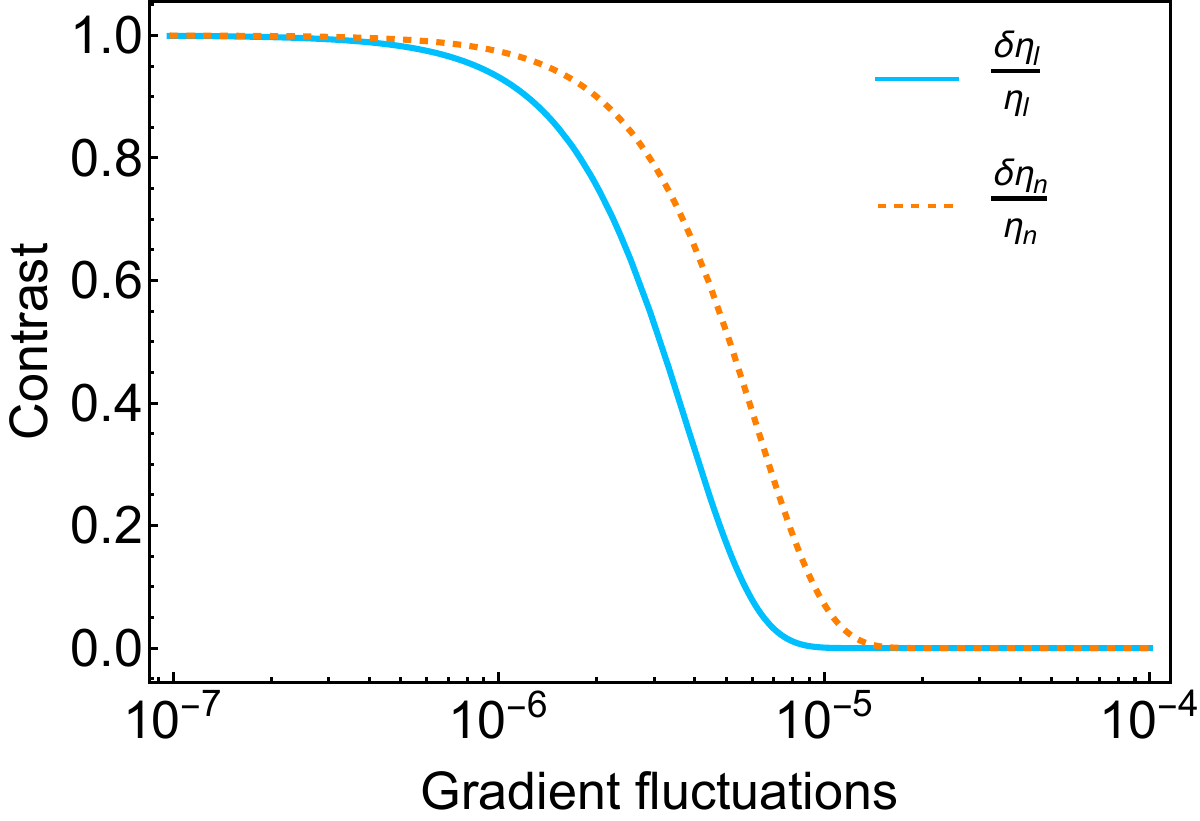}
		\caption{Wave packet contrast as a function of gradient fluctuations. The solid blue line represents contrast variation due to fluctuations in the linear magnetic field gradient, while the orange dashed line corresponds to fluctuations in the nonlinear magnetic field gradient. The horizontal axis shows the dimensionless gradient fluctuation normalized by the corresponding gradient. Here we take the maximum value of the gradient used in the scheme, $\eta_{l}=5\times 10^{3}\,\text{T/m}$, $\eta_{n}=1\times 10^{6}\,\text{T}/\text{m}^{2}$. Other parameters include mass $m = 10^{-15}$ kg and initial wave packet width $\sigma_{0} = 2\times 10^{-11}$ m, with additional values listed in Table~\ref{values_of_parameters}.
		\label{contrast_vs_gradient_fluctuations}}
	\end{figure}

	To calculate the contrast, we also need the wave packet width at the final stage. Since analytical expressions exist for wave packet evolution in HP and IHP (Eqs.~(\ref{width_in_HP2}) and (\ref{width_in_IHP2})), we iteratively substitute initial conditions and propagate the wave packet width through each stage until the fifth stage is reached. This process is solved numerically, and the resulting contrast variation with linear gradient fluctuations is shown in Fig.~\ref{contrast_vs_gradient_fluctuations}. 
	The procedure for evaluating the effect of nonlinear magnetic field gradient fluctuations on contrast is similar. Here, we assume a fluctuation $\delta\eta_{n}$ in the nonlinear magnetic field gradient while keeping the linear gradient constant. Again, by combining the equations of motion from stage 1 to stage 5, we compute the corresponding position and momentum deviations. Substituting these into Eq.~(\ref{contrast}) and calculating the resulting wave packet width evolution, we obtain the contrast variation as a function of $\delta\eta_{n}$, shown in Fig.~\ref{contrast_vs_gradient_fluctuations}.  
	
	From Fig.~\ref{contrast_vs_gradient_fluctuations}, we observe that to maintain 99\% contrast, the gradient fluctuation must be below $10^{-7}$ for both linear and nonlinear magnetic fields. Interestingly, nonlinear magnetic fields have also been used to generate a macroscopic spatial superposition state in \cite{Zhou:2022frl}, but without employing IHP. In that case, to maintain 99\% contrast, the gradient fluctuation must be below $10^{-9}$. This means that incorporating IHP relaxes the constraint on nonlinear gradient fluctuations by two orders of magnitude. This improvement arises because IHP increases the wave packet width, making the system more tolerant to position and momentum deviations caused by gradient fluctuations.

	\subsection{Deviation in initial position and wave packet contrast}
\begin{figure}[htp]
		\centering
		\includegraphics[width=\linewidth]{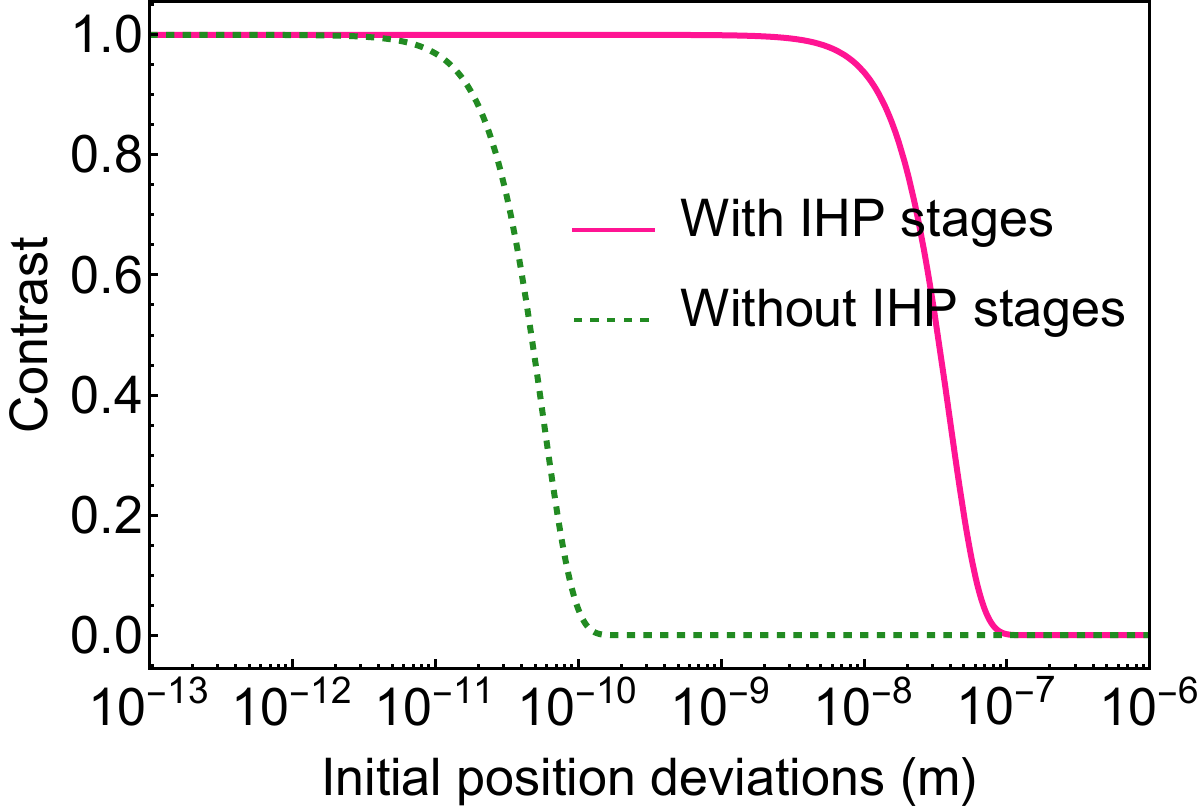}
		 \caption{Wave packet contrast as a function of initial position deviations. The solid pink line represents the contrast when the wave packets experienced the IHP stages (the contrast of the final stage). The green dashed line represents the contrast that the wave packets did not experience the IHP stages. At this point, only the first stage is considered and the wave packets are allowed to recombine in the first stage. We take the values of mass $m = 10^{-15}$ kg and initial wave packet width $\sigma_{0} = 2\times 10^{-11}$ m, and other values listed in Table~\ref{values_of_parameters}.\label{contrast_with_IPD}}
	\end{figure}

	In this section, we examine how initial position deviations affect the wave packet contrast, given by Eq.~(\ref{contrast}). Ideally, the classical trajectory is perfectly closed, resulting in $\Delta x = 0$ and $\Delta b = 0$. However, small deviations in the initial position introduce changes in $\Delta x$ and $\Delta b$, altering the final contrast. Here, we define $\Delta x$ and $\Delta b$ as:
	\begin{align}\label{diviation3}
		\Delta x &= x(x_{0} + \delta x, t) - x(x_{0}, t), \nonumber\\
		\Delta b &= b(p(x_{0} + \delta x), t) - b(p(x_{0}), t),
	\end{align}
	where $\delta x$ is a small constant representing the deviation from the initial position. To compare the effect of initial position deviation on contrast in different scenarios, we first analyze the case where only the HP stage is considered, without the influence of the IHP stages. We assume that the wave packet completes a full period in the first HP stage and that the two wave packets recombine at the end of this stage. By substituting the equations of motion for the HP case into Eq.(\ref{diviation3}) and combining them with Eq.(\ref{contrast}), we obtain the contrast as a function of the initial position deviation:
	\begin{align}\label{contrast2}
		C = \exp\left(-\frac{\delta x^{2}}{8\sigma_{0}^{2}}\right).
	\end{align}
	Here, the contrast depends only on the initial position deviation and the initial wave packet width $\sigma_0$. The relationship between contrast and $\delta x$ given in Eq.(\ref{contrast2}) is shown in Fig.\ref{contrast_with_IPD}. In the HP-only case, to maintain a contrast greater than 99\%, the initial position deviation must be within the width of the initial wave packet, which is approximately $10^{-11}$ m.
	
	Next, we examine how initial position deviations influence the final contrast when the IHP stages are included. The calculation follows a similar approach to that used for analyzing the effect of gradient fluctuations on contrast. Using the equations of motion from stages 1 to 5, we determine the position and momentum deviations caused by the initial position deviation. The wave packet width is then computed numerically by incorporating the wave packet evolution equations for both HP and IHP stages. Finally, by substituting the computed position and momentum deviations, as well as the wave packet width, into Eq.(\ref{contrast}), we obtain the contrast as a function of initial position deviation, illustrated in Fig.\ref{contrast_with_IPD}.
	
	From Fig.\ref{contrast_with_IPD}, we observe that when the IHP stages are included, the initial position deviation can be as large as $10^{-9}$ m while still maintaining a contrast above 99\%. This suggests that although the IHP stage amplifies the effect of initial position deviations, its impact on increasing the wave packet width is even more significant. As a result, the requirement for initial position accuracy is relaxed by two orders of magnitude while maintaining the same contrast level, compared to the case without the IHP stages.

	\section{Conclusion}\label{conclusion}
	
	In this paper, we explored the use of spin-dependent forces and IHP to generate large spatial superposition states of massive objects and construct a full-loop interferometer. The spin-dependent force enables a small initial spatial separation of the massive wave packet within a short time, while the IHP subsequently amplifies this separation significantly. For a nanodiamond with a mass of $10^{-15}$ kg, our analysis shows that a superposition size of approximately 50 $\mu$m can be achieved within 0.1 s. We also provided an analytical treatment of wave packet evolution in both the HP and IHP using path integral methods.
	
	Based on the equations of motion for each stage and the exact wave packet evolution, we analyzed the impact of gradient fluctuations and initial position deviations on the interference contrast. Our results indicate that incorporating the IHP significantly relaxes the experimental constraints. Specifically, the tolerance for nonlinear magnetic field gradient fluctuations improves by two orders of magnitude, from $10^{-9}$ to $10^{-7}$, while maintaining 99\% contrast. Similarly, the requirement for position control accuracy is reduced by two orders of magnitude, from $10^{-11}$ m to $10^{-9}$ m, when considering both the HP and IHP stages. This relaxation arises because the IHP enhances wave packet expansion more effectively than it amplifies gradient fluctuations and position deviations, thereby increasing robustness against these imperfections.
	
	Our study focused on a nanodiamond embedded with an NV centre. The nanodiamond possesses mechanical degrees of freedom (CoM motion), internal degrees of freedom (phonons), and rotational degrees of freedom. Currently, the CoM motion can be cooled to the ground state \cite{Delic:2020ndp}, and the internal phonons are difficult to excite during movement in HP and IHP \cite{Henkel:2021wmj,Xiang:2024zol}. The rotational degree of freedom can affect the final wave packet contrast of the interferometer; however, if the NV center is at or near the nanodiamond center, this effect can be mitigated by fine-tuning the magnetic field and timing \cite{Japha:2022xyg,Japha:2022phw,Zhou:2024pdl}. Furthermore, the electron spin coherence time of the NV center can be as long as 1 s at low temperatures and under pure nanocrystal conditions \cite{Bar-Gill2013,Frangeskou:2016mqs,Abobeih2018}. These findings indicate the feasibility of realizing macroscopic quantum states with masses ranging from $10^{-17}$ to $10^{-14}$ kg and superposition sizes on the order of micrometres in the laboratory.
	
	\textbf{Note added:} Recent related independent work \cite{Braccini:2024fey} has come to our attention.
	
	\section*{acknowledgments} We are grateful to Tian Zhou, Ryan Marshman, Ryan Rizaldy and Sougato Bose for stimulating discussions. R. Z. and Q. X. are supported by China Scholarship Council (CSC) fellowship. AM's research is funded in part by the Gordon and Betty Moore Foundation through Grant GBMF12328, DOI 10.37807/GBMF12328. This material is based upon work supported by Alfred P. Sloan Foundation under Grant No. G-2023-21130.
	
	\bibliographystyle{unsrt}
	\bibliography{references}
	\onecolumngrid
	\begin{appendices}
\section{One dimensional motion approximation}\label{one_dimension_motion}
\begin{figure*}[htp]
	\centering
	\begin{subfigure}[b]{0.33\textwidth}
		\centering
		\includegraphics[width=\textwidth]{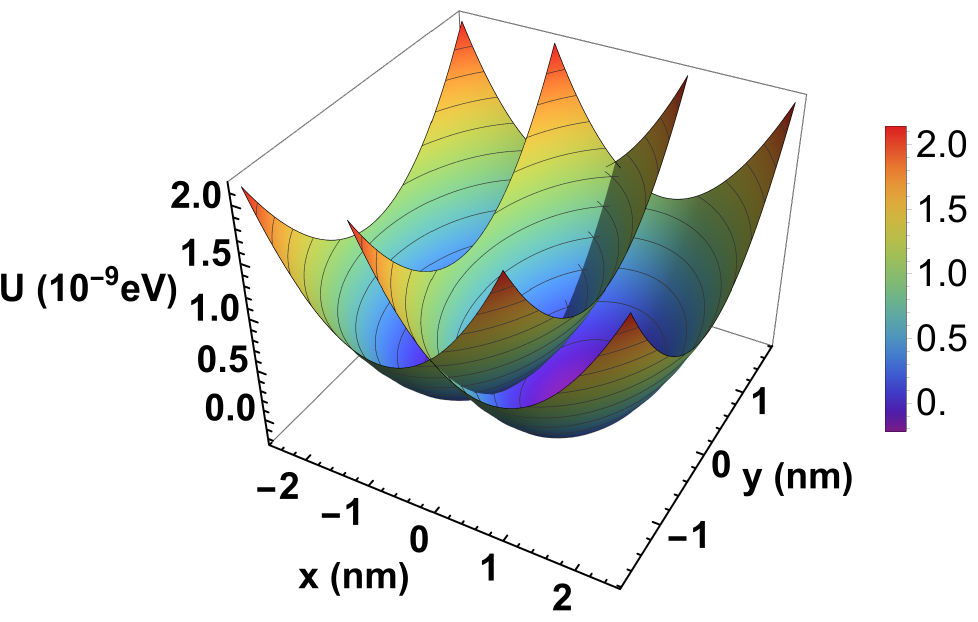}
		\caption{ }
		\label{HP3D}
	\end{subfigure}
	\hfil
	\begin{subfigure}[b]{0.3\textwidth}
		\centering
		\includegraphics[width=\textwidth]{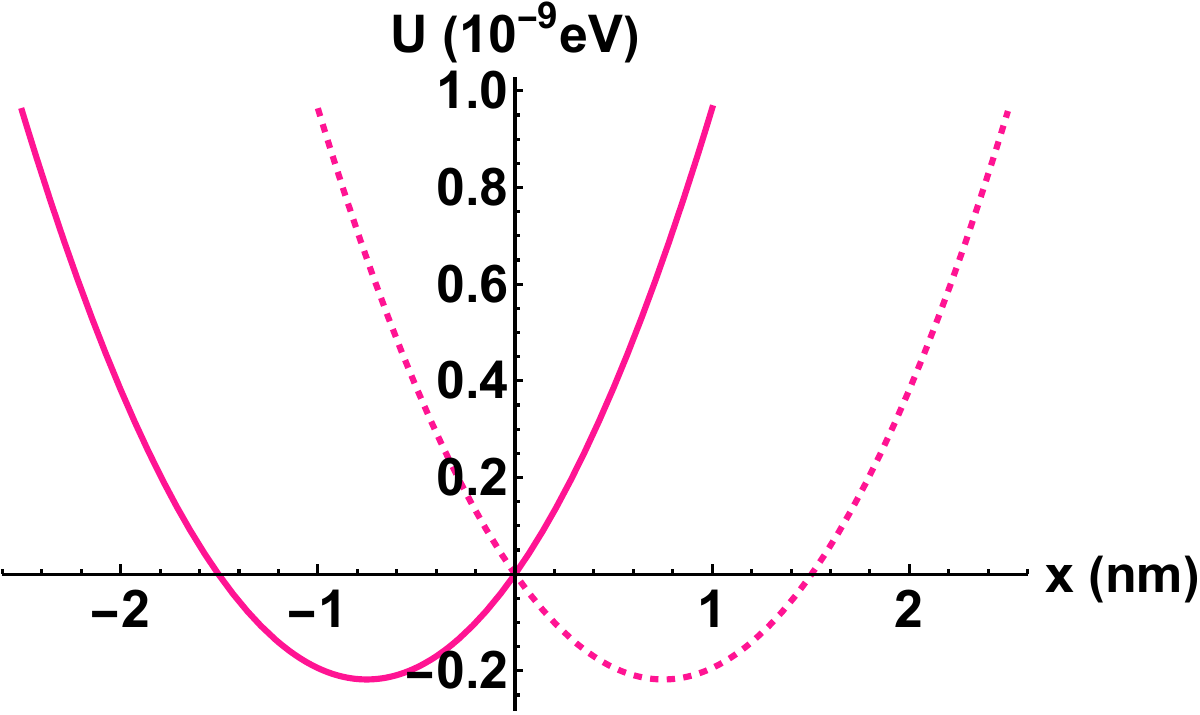}
		\caption{ }
		\label{HP_x_direction}
	\end{subfigure}
	\hfil
	\begin{subfigure}[b]{0.3\textwidth}
		\centering
		\includegraphics[width=\textwidth]{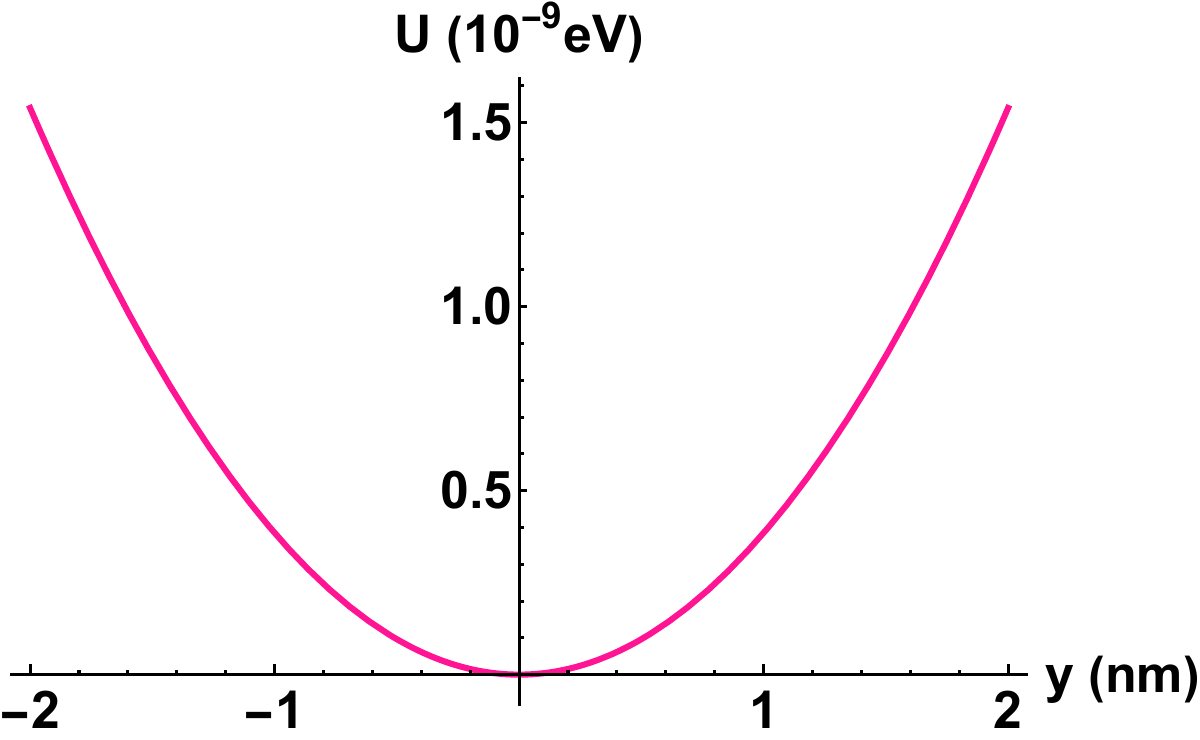}
		\caption{ }
		\label{HP_y_direction}
	\end{subfigure}
	\\[0.1cm]
	\centering
	\begin{subfigure}[b]{0.33\textwidth}
		\centering
		\includegraphics[width=\textwidth]{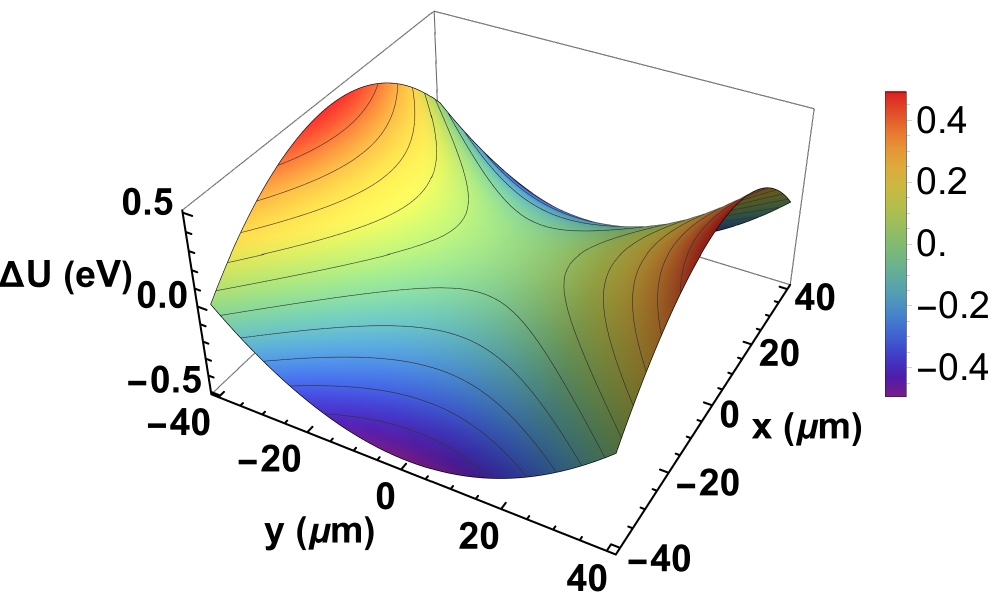}
		\caption{ }
		\label{IHP3D}
	\end{subfigure}
	\hfil
	\begin{subfigure}[b]{0.3\textwidth}
		\centering
		\includegraphics[width=\textwidth]{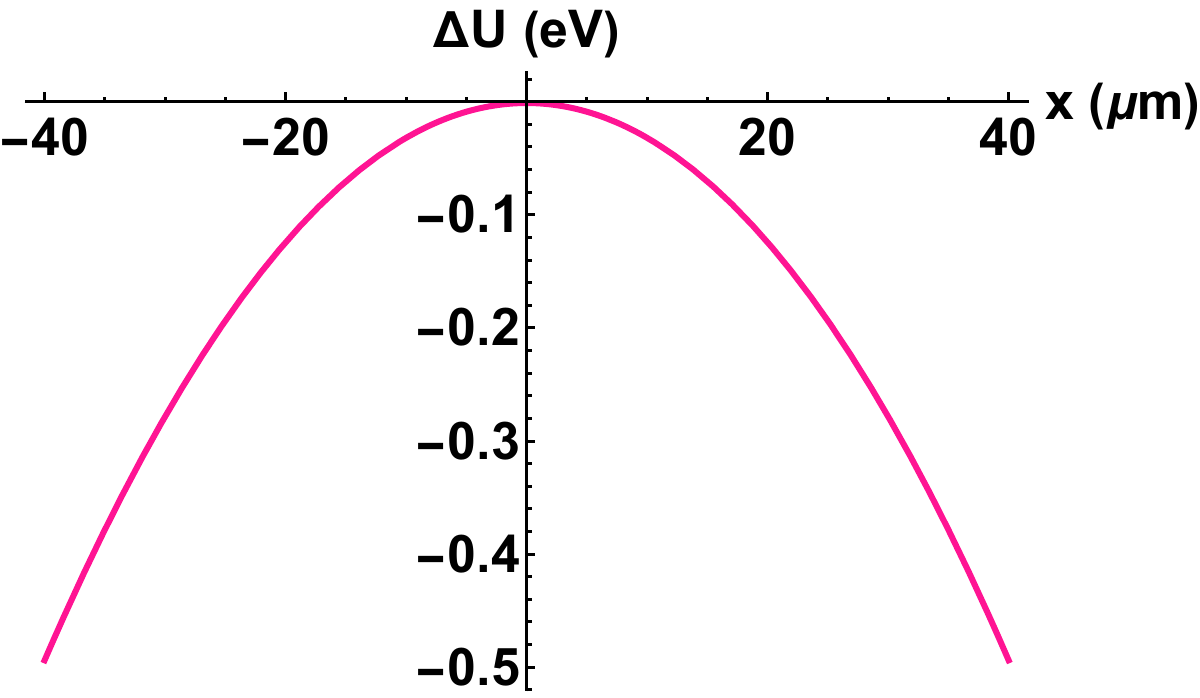}
		\caption{ }
		\label{IHP_x_direction}
	\end{subfigure}
	\hfil
	\begin{subfigure}[b]{0.3\textwidth}
		\centering
		\includegraphics[width=\textwidth]{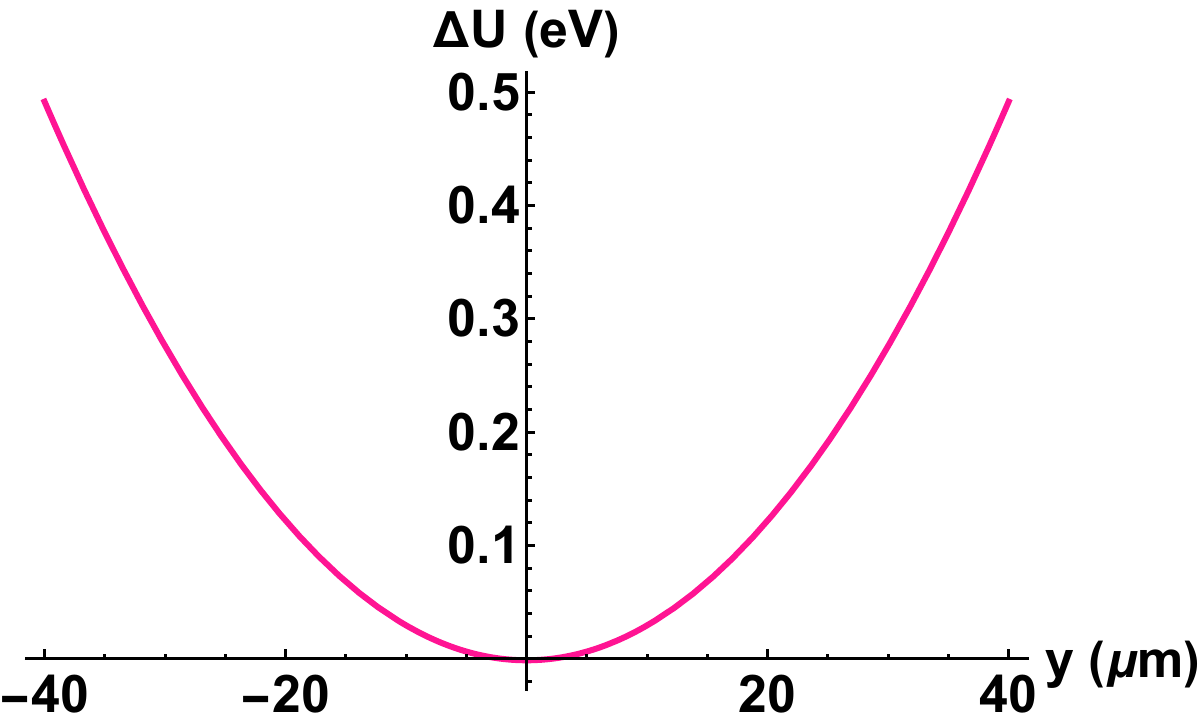}
		\caption{ }
		\label{IHP_y_direction}
	\end{subfigure}
	\caption{Comparison of the potential landscapes for the linear and nonlinear magnetic fields. (a) Three-dimensional plot of the potential corresponding to the linear magnetic field. (b) Potential profile in the $x$-direction for the linear magnetic field when $y=0$. The solid pink line corresponds to $S_{x} = 1$, while the dashed pink line corresponds to $S_{x} = -1$. (c) Potential profile in the $y$-direction for the linear magnetic field when $x=0$. Since $x=0$, the potentials for spin states $S_{x} = \pm 1$ coincide in the $y$-direction. (d) Three-dimensional plot of the potential corresponding to the nonlinear magnetic field. (e) Potential profile in the $x$-direction for the nonlinear magnetic field when $y=0$. (f) Potential profile in the $y$-direction for the nonlinear magnetic field when $x=0$. For visualization purposes, we set $B_{0} = 0$ in the linear magnetic field case. In the nonlinear case, we introduce a reference potential $U_{0} = 1539.84$ eV and define $\Delta U = U - U_{0}$, ensuring that the potential variation starts from zero, as shown in (e) and (f). The values of other physical parameters are listed in Table~\ref{values_of_parameters2}.}
	\label{comparison_of_potentials}
\end{figure*}

For a static magnetic field, Maxwell's equations require that  $\div{\vb{B}}=0$ and $\curl{\vb{B}}=0$. In our scheme, the complete linear magnetic field is written as
\begin{align}\label{complete_linear_field}
	\vb{B}=(B_{0}+\eta_{l}x)\vb{e}_{x} - \eta_{l}y\vb{e}_{y},
\end{align} 
where $B_{0}$ is the bias field along the $x$-direction, $\eta_{l}$ is the gradient of the linear field (identical in the $x$ and $y$ directions), and $\vb{e}_{x}$, $\vb{e}_{y}$ are the unit vectors in the $x$ and $y$ directions, respectively. Similarly, the complete nonlinear magnetic field is given by
\begin{align}\label{complete_nonlinear_field}
	\vb{B}=(B_{0} - \eta_{n}x^{2} + \eta_{n}y^{2})\vb{e}_{x} - 2\eta_{n}xy\vb{e}_{y},
\end{align}
where $\eta_{n}$ is the nonlinear gradient with unit T/m$^2$. Other symbols have the same meanings as in the linear case. The potential energy for a nanodiamond embedded with an NV center subjected to an external magnetic field is
\begin{align}\label{potential}
	U=-\frac{\chi_{\rho}m}{2\mu_{0}}\vb{B}^{2}+\hbar\gamma_{e}\vb{S}\vdot\vb{B},
\end{align}
where $\chi_{\rho}$ is the mass susceptibility of the diamond, $m$ is its mass, $\mu_{0}$ is the vacuum permeability, $\hbar$ is the reduced Planck constant, $\gamma_{e}$ is the electron gyromagnetic ratio, and $\vb{S}$ is the spin of the NV center. By substituting Eqs.~(\ref{complete_linear_field}) and (\ref{complete_nonlinear_field}) into Eq.~(\ref{potential}), we obtain the potential energy corresponding to both linear and nonlinear magnetic fields, as illustrated in Fig.~\ref{comparison_of_potentials}.

\subsection*{Force in the Linear Magnetic Field}

Substituting Eq.~\eqref{complete_linear_field} into Eq.~\eqref{potential}, the force in the linear magnetic field is
\begin{align}\label{linear_force}
	\vb{F}&=-\grad U_{l},\nonumber\\
	&=\qty(\frac{\chi_{\rho}m}{\mu_{0}}\eta_{l}^{2}x + \frac{\chi_{\rho}m}{\mu_{0}}B_{0}\eta_{l} - \hbar\gamma_{e}S_{x}\eta_{l})\vb{e}_{x} + \qty(\frac{\chi_{\rho}m}{\mu_{0}}\eta_{l}^{2}y + \hbar\gamma_{e}S_{y}\eta_{l})\vb{e}_{y},
\end{align}
where $U_{l}$ is the potential energy associated with the linear field. Table~\ref{values_of_parameters2} lists typical values of the physical constants. Advances in current technology have enabled the cooling of nanoparticles with masses around $10^{-18}$~kg to their motional ground states~\cite{Piotrowski:2022qda, Delic:2020ndp}, and by further improving the environmental conditions, even larger masses (e.g. $10^{-15}$~kg) can be cooled \cite{TebbenjohannsEtAl2021}. Notice that in Eq.~\eqref{linear_force} the force has a $y$-component. However, because a bias magnetic field $B_{0}$ is applied along $x$, the electron spin undergoes Larmor precession about the $x$-axis with frequency $\omega_{L}=\abs{\gamma_{e}B_{0}}\approx 10^{12}$ Hz, which is much faster than the experimental frequency (approximately 10 Hz). Hence, the average value of $S_{y}$ is zero. Furthermore, the restoring force from the $y$-component creates a simple harmonic motion with an amplitude of about $10^{-11}$~m (the width of the ground state wave packet), which is negligible compared to the superposition size we aim to achieve. In addition, we can further restrict the $y$ motion by applying an external trapping potential. For instance, Ref.~\cite{Elahi:2024dbb} employs a three-dimensional magnetic field generated by a current-carrying wire that levitates nanodiamonds in the $z$ direction, confines motion in $y$, and allows free evolution in $x$. Therefore, it is reasonable to consider the nanoparticle motion as one-dimensional (along $x$).

\begin{table}[t]
	\centering
	\setlength{\tabcolsep}{-0.1\tabcolsep}
	\renewcommand{\arraystretch}{2}
	\begin{tabular}{p{4cm} p{2cm} p{2.5cm} p{2.5cm} p{2.5cm} p{2.5cm}}
		\hline
		$\chi_{\rho}m/\mu_{0}$ $(\text{m}^{2}\text{s}^{2}\text{A}^{2} \text{kg}^{-1})$ &$B_{0}$ $(\text{T})$ 
		&$\hbar$ $(\text{kg}\text{m}^{2}\text{s}^{-1})$ 
		&$\gamma_{e}$ $(\text{C}\text{kg}^{-1})$
		&$\eta_{l}$ $(\text{T}\text{m}^{-1})$ 
		&$\eta_{n}$ $(\text{T}\text{m}^{-2})$ \\ \hline
		$\approx5\times10^{-18}$
		&10
		&$\approx 1\times 10^{-34}$
		&$\approx 2\times 10^{11}$
		&$\approx 5\times 10^{3}$
		&$\approx 1\times 10^{6}$\\ \hline
	\end{tabular}
	\caption{Values of the physical parameters used for force estimation. Here $B_{0}$ is the bias magnetic field, $\eta_{l}$ and $\eta_{n}$ are the gradients of the linear and nonlinear magnetic fields, respectively. The mass $m$ is taken as $10^{-15}$ kg. $\chi_{\rho}$, $\mu_{0}$, $\hbar$, and $\gamma_{e}$ are fundamental constants.}
	\label{values_of_parameters2}
\end{table}

\subsection*{Force in the Nonlinear Magnetic Field}

By substituting Eq.~\eqref{complete_nonlinear_field} into Eq.~\eqref{potential}, the force in the nonlinear magnetic field is given by 
\begin{align}\label{nonlinear_force}
	\vb{F}&=-\grad U_{nl},\nonumber\\
	&=\qty(\frac{2\chi_{\rho}m}{\mu_{0}}\eta_{n}^{2}x^{3} - \frac{2\chi_{\rho}m}{\mu_{0}}B_{0}\eta_{n}x 
	+ 2\hbar\gamma_{e}S_{x}\eta_{n}x 
	+ \frac{2\chi_{\rho}m}{\mu_{0}}\eta_{n}^{2}y^{2}x + 2\hbar\gamma_{e}S_{y}\eta_{n}y )\vb{e}_{x}\nonumber\\
	&\quad +\qty(\frac{2\chi_{\rho}m}{\mu_{0}}\eta_{n}^{2}y^{3} 
	+ \frac{2\chi_{\rho}m}{\mu_{0}}B_{0}\eta_{n}y 
	- 2\hbar\gamma_{e}S_{x}\eta_{n}y 
	+ \frac{2\chi_{\rho}m}{\mu_{0}}\eta_{n}^{2}x^{2}y 
	+ 2\hbar\gamma_{e}S_{y}\eta_{n}y)\vb{e}_{y},
\end{align}
where $U_{nl}$ is the potential associated with the nonlinear field.  During the nonlinear field stage, the spin states transform to $\ket{S_{x}} = \ket{0}$ and $\ket{S_{y}} = (\ket{+1}+\ket{-1})/\sqrt{2}$. Due to the bias magnetic field $B_{0}$ along $x$, the $S_{y}$ component undergoes rapid Larmor precession, leading to an average value of zero. Consequently, the terms involving $S_{x}$ and $S_{y}$ in Eq.~(\ref{nonlinear_force}) can be neglected. Substituting parameter values from Table~\ref{values_of_parameters2} into Eq.~\eqref{nonlinear_force}\footnote{Here, $x$ is taken up to 25~$\mu$m, considering that the maximum superposition size is around 50~$\mu$m.}, we find that the dominant term in the $x$-direction is $(-2\chi_{\rho}m/\mu_{0})B_{0}\eta_{n}x$. This term corresponds to the IHP, as shown in Fig.~\ref{IHP_x_direction}, and is responsible for the rapid separation of the wave packets. Similarly, the dominant term in the $y$-direction is $(-2\chi_{\rho}m/\mu_{0})B_{0}\eta_{n}y$,which corresponds to HP, as illustrated in Fig.~\ref{IHP_y_direction}. This term effectively restricts motion in the $y$-direction. If the initial state is the ground state, then the motion of the wave packet in the $y$-direction follows simple harmonic oscillation with an amplitude comparable to the width of the initial wave packet (approximately $10^{-11}$~m). Since this amplitude is negligible in comparison to the superposition size, the motion along $y$ can be effectively disregarded.
As a result, during the nonlinear field stage, the nanoparticle's motion remains predominantly confined to the $x$-direction. The effective potential in this regime ensures rapid separation of the wave packets along $x$, while simultaneously suppressing displacement along $y$.

In summary, both for linear and nonlinear magnetic field configurations, the nanoparticle experiences forces with components in both $x$ and $y$. However, due to the rapid Larmor precession along $x$ (causing the mean $S_{y}$ to vanish) and an effective trapping potential that confines $y$-motion, we can safely approximate the nanoparticle motion as one-dimensional along the $x$-axis.

	\section{Wave packet evolution}\label{path_integral}
	According to the path integral, the evolution of the wave function can be written as:
	\begin{align}\label{wave_function_evolution}
		\psi(x,t)=\int dx' K(x,t;x',0)\psi(x',0),
	\end{align}
	where $\psi(x',0)$ represents the wave function at the initial moment. $K(x,t;x',0)$ is the propagator. When the potential energy is quadratic, the propagator can be calculated by the Van Vleck-Pauli-Morette formula:
	\begin{align}\label{propagator}
		K(x_{f},t_{f};x_{i},t_{i})=\sqrt{\frac{i}{2\pi \hbar}\pdv{S}{x_{f}}{x_{i}}}\exp[\frac{i}{\hbar}S],
	\end{align}
	where $S_{c}$ is the classical action quantity and defined as:
	\begin{align}\label{classical_action}
		S=\int_{t_{i}}^{t_{f}}dt\, \mathcal{L}(t).
	\end{align}
	$\mathcal{L}$(t) is the Lagrangian of the system. Assuming that the solution to the classical trajectory of the system is $x_{c}(t)$, the Lagrangian can be written as: 
	\begin{align}\label{lagrangian}
		\mathcal{L}(t)=\frac{1}{2}m\dot{x}^{2}(t)-\frac{1}{2}m\omega^{2}x^{2}(t).
	\end{align}
	\subsection{Wave packet evolution in a HP}
	The general solution for the classical trajectory of a wave packet in a HP is:
	\begin{align}\label{classical_solution_in_a_HP1}
		x(t)=x_{0}\cos(\omega t)+\frac{p_{0}}{m\omega}\sin(\omega t),
	\end{align}
	where $x_{0}$ and $p_{0}$ are the classical initial position and initial momentum of the wave packet, respectively. Consider the boundary conditions $x(t_{i}=0)=x_{i}$ and $x(t_{f})=x_{f}$. Substituting them into Eq.(\ref{classical_solution_in_a_HP1}) yields:
	\begin{align}\label{classical_solution_in_a_HP2}
		x(t)=x_{i}\cos(\omega t)+\frac{x_{f}-x_{i}\cos(\omega t_{f})}{\sin(\omega t_{f})}\sin(\omega t).
	\end{align}
	Combining Eq.(\ref{classical_action}), (\ref{lagrangian}) and (\ref{classical_solution_in_a_HP2}) gives the classical action at the harmonic  potential as:
	\begin{align}\label{action_in_a_HP}
		S=\frac{m\omega}{2}\frac{(x_{f}^{2}+x_{i}^{2})\cos(\omega t)-2x_{f}x_{i}}{\sin(\omega t)}.
	\end{align}
	Note that after integrating in Eq.(\ref{classical_action}), the time parameter in the action is ``$t_{f}$''. In Eq.(\ref{action_in_a_HP}) we replace ``$t_{f}$'' with ``$t$'', thus aligning with the time variable in Eq.(\ref{wave_function_evolution}). Substituting Eq.(\ref{action_in_a_HP}) into Eq(\ref{propagator}) results in the propagator of the wave packet at the HP as:
	\begin{align}\label{propagator_for_HP}
		K(x_{f},t;x_{i},0)=\sqrt{\frac{m\omega}{i2\pi\hbar\sin(\omega t)}}\exp[\frac{i}{\hbar}\frac{m\omega}{2}\frac{(x_{f}^{2}+x_{i}^{2})\cos(\omega t)-2x_{f}x_{i}}{\sin(\omega t)}].
	\end{align}
	Since both the initial wave function (Eq.(\ref{initial_wave_function})) and the propagator (Eq.(\ref{propagator_for_HP})) are Gaussian quadratic functions, solving Eq.(\ref{wave_function_evolution}) for the wave packet evolution is a Gaussian quadratic integral. The result of the integration is still a Gaussian quadratic function:
	\begin{align}\label{evolution_of_WF_in_HP}
		\psi(x,t)=N(t)\exp[i\frac{x^{2}}{4u_t^{2}}-\frac{x_{0}^{2}}{4\sigma_{0}^{2}}]\exp[\frac{\qty(ib_{0}-i\frac{x}{2u_{t}^{2}\cos(\omega t)}+\frac{x_{0}}{2\sigma_{0}^{2}})^{2}}{\frac{1}{\sigma_{0}^{2}}-i\qty(\frac{1}{u_{t}^{2}}+a_{0})}],
	\end{align}
	where 
	\begin{align}
		u_{t}^{2}&=\frac{\hbar\sin(\omega t)}{2m\omega\cos(\omega t)},\nonumber\\
		N(t)&=N_{0}\sqrt{\frac{m\omega}{i2\pi\hbar\sin(\omega t)}}\sqrt{\frac{4\pi}{1/\sigma_{0}^{2}-i\qty(1/u_{t}^{2}+a_{0})}}e^{ic_{0}}.
	\end{align}
	Eq.(\ref{evolution_of_WF_in_HP}) can be rewritten in the familiar form of the GSWP:
	\begin{align}\label{solution_for_HP}
		\psi(x,t)=N(t)\exp[-\frac{1}{4\sigma_{x}^{2}(t)}\qty(x-x_{c}(t))^{2}+i\qty(\frac{a(t)}{4}x^{2}+b(t)x+c(t))],
	\end{align}
	where
	\begin{align}
		\sigma_{x}(t)&=\sigma_{0}\qty(\frac{\hbar^{2}}{4m^{2}\omega^{2}\sigma_{0}^{4}}\sin^{2}(\omega t)+\qty(\frac{\hbar a_{0} }{2m\omega}\sin(\omega t)+\cos(\omega t))^{2})^{\frac{1}{2}},\\
		x_{c}(t)&=\frac{\hbar a_{0} x_{0}}{2m\omega}\sin(\omega t)+x_{0}\cos(\omega t)+\frac{\hbar b_{0} }{m\omega}\sin(\omega t),
	\end{align}
	which represent the spatial width of the wave packet and the classical equation of motion of the nanoparticle, respectively. The expressions for the three parameters in the imaginary part are:
	\begin{align}\label{parameters_abc_in_HP}
		a(t)&=\frac{1}{u_{t}^{2}}
		-\frac{1+a_{0} u_{t}^{2}}{4u_{t}^{6}\cos^{2}(\omega t)\qty(\qty(1/u_{t}^{2}+a_{0})^{2}+1/\sigma_{0}^{4})},\nonumber\\
		b(t)&=\frac{2b_{0}\sigma_{0}^{4}-u_{t}^{2}\qty(x_{0}-2a_{0}b_{0}\sigma_{0}^{4})}{2\cos(\omega t)\qty(\sigma_{0}^{4}+2a_{0}u_{t}^{2}\sigma_{0}^{4}+u_{t}^{4}\qty(1+a_{0}^{2}\sigma_{0}^{4}))},\nonumber\\
		c(t)&=\frac{x_{0}^{2}+x_{0}u_{t}^{2}(4b_{0}+a_{0}x_{0})-4b_{0}^{2}\sigma_{0}^{4}(1+a_{0}u_{t}^{2})}{4u_{t}^{2}\sigma_{0}^{4}\qty(\qty(1/u_{t}^{2}+a_{0})^{2}+1/\sigma_{0}^{4})}.
	\end{align}
	
	\subsection{Wave packet evolution in a IHP}
	The calculation process for the evolution of the wave packet in the IHP is the same as in the case of the HP. The form of the classical equation of motion and the action of the wave packet in the IHP are the same as in the HP case, but with the replacement of $``\sin"$ with $``\sinh"$ and $``\cos"$ with $``\cosh"$. According to Eq.(\ref{propagator}), the propagator at the IHP is obtained as:
	\begin{align}\label{propagator_for_IHP}
		K'(x_{f},t;x_{i},0)=\sqrt{\frac{m\omega}{i2\pi\hbar\sinh(\omega t)}}\exp[\frac{i}{\hbar}\frac{m\omega}{2}\frac{(x_{f}^{2}+x_{i}^{2})\cosh(\omega t)-2x_{f}x_{i}}{\sinh(\omega t)}].
	\end{align}
	Using Eq.(\ref{wave_function_evolution}) again, multiplying this propagator with the initial wave function and integrating over the initial position gives:
	\begin{align}\label{evolution_of_WF_in_IHP}
		\psi'(x,t)=N'(t)\exp[i\frac{x^{2}}{4v_t^{2}}-\frac{x_{0}^{2}}{4\sigma_{0}^{2}}]\exp[\frac{\qty(ib_{0}-i\frac{x}{2v_{t}^{2}\cosh(\omega t)}+\frac{x_{0}}{2\sigma_{0}^{2}})^{2}}{\frac{1}{\sigma_{0}^{2}}-i\qty(\frac{1}{v_{t}^{2}}+a_{0})}],
	\end{align}
	where 
	\begin{align}
		v_{t}^{2}&=\frac{\hbar\sinh(\omega t)}{2m\omega\cosh(\omega t)},\nonumber\\
		N'(t)&=N_{0}\sqrt{\frac{m\omega}{i2\pi\hbar\sinh(\omega t)}}\sqrt{\frac{4\pi}{1/\sigma_{0}^{2}-i\qty(1/v_{t}^{2}+a_{0})}}e^{ic_{0}}.
	\end{align}
	Rearranging Eq.(\ref{evolution_of_WF_in_IHP}) yields:
	\begin{align}\label{solution_for_IHP}
		\psi'(x,t)=N'(t)\exp[-\frac{1}{4\sigma_{x}'^{2}(t)}\qty(x-x'_{c}(t))^{2}+i\qty(\frac{a'(t)}{4}x^{2}+b'(t)x+c'(t))],
	\end{align}
	where
	\begin{align}
		\sigma_{x}'(t)&=\sigma_{0}\qty(\frac{\hbar^{2}}{4m^{2}\omega^{2}\sigma_{0}^{4}}\sinh^{2}(\omega t)+\qty(\frac{\hbar a_{0} }{2m\omega}\sinh(\omega t)+\cosh(\omega t))^{2})^{\frac{1}{2}},\\
		x'_{c}(t)&=\frac{\hbar a_{0} x_{0}}{2m\omega}\sinh(\omega t)+x_{0}\cosh(\omega t)+\frac{\hbar b_{0} }{m\omega}\sinh(\omega t),
	\end{align}
	which are the spatial width of the wave packet and the classical equation of motion at the IHP. The expressions for the parameters of the imaginary part are:
	\begin{align}\label{parameters_abc_in_IHP}
		a'(t)&=\frac{1}{v_{t}^{2}}
		-\frac{1+a_{0} v_{t}^{2}}{4v_{t}^{6}\cosh^{2}(\omega t)\qty(\qty(1/v_{t}^{2}+a_{0})^{2}+1/\sigma_{0}^{4})},\nonumber\\
		b'(t)&=\frac{2b_{0}\sigma_{0}^{4}-v_{t}^{2}\qty(x_{0}-2a_{0}b_{0}\sigma_{0}^{4})}{2\cosh(\omega t)\qty(\sigma_{0}^{4}+2a_{0}v_{t}^{2}\sigma_{0}^{4}+v_{t}^{4}\qty(1+a_{0}^{2}\sigma_{0}^{4}))},\nonumber\\
		c'(t)&=\frac{x_{0}^{2}+x_{0}v_{t}^{2}(4b_{0}+a_{0}x_{0})-4b_{0}^{2}\sigma_{0}^{4}(1+a_{0}v_{t}^{2})}{4v_{t}^{2}\sigma_{0}^{4}\qty(\qty(1/v_{t}^{2}+a_{0})^{2}+1/\sigma_{0}^{4})}.
	\end{align}

\end{appendices}
	
\end{document}